\documentclass[twocolumn]{aastex631}

\shorttitle{Recurrent Nova U Scorpii is Not a Supernova Progenitor}
\shortauthors{Schaefer}
\graphicspath{{./}{figures/}}

\begin{document}
\title{Orbital Period Changes of Recurrent Nova U Scorpii Demonstrate that M$_{\rm ejecta}$=26$\times$M$_{\rm accreted}$ and Is Not a Type Ia Supernova Progenitor}

\author[0000-0002-2659-8763]{Bradley E. Schaefer}
\affiliation{Department of Physics and Astronomy,
Louisiana State University, Baton Rouge, LA 70803, USA}

\author[0000-0002-9810-0506]{Gordon Myers}
\affiliation{AAVSO, Inverness Way, Hillsborough, CA 94010, USA; }

\begin{abstract}

Recurrent nova U Scorpii (U Sco) is one of the prototypes for a Type Ia supernova progenitor.  The logic is that the white dwarf is near the Chandrasekhar mass and gas is accumulating onto its surface at a near-maximal accretion rate, so it will soon increase its mass to the supernova trigger.  But the white dwarf loses mass every nova eruption, so the issue is balancing the mass ejected ($M_{\rm ejecta}$) against the mass accreted between eruptions ($M_{\rm accreted}$).  Measuring $M_{\rm accreted}$ can be done in several ways to useable accuracy.  But the old methods for measuring $M_{\rm ejecta}$ (involving the flux in hydrogen emission lines) are all with real error bars of 2--3 orders of magnitude.  The only solution is to measure the change of the orbital period across the nova eruption ($\Delta P$).  But this solution requires a vast photometric program of eclipse timings stretching decades.  For U Sco, a program started in 1989, now reaches its culmination with measures of $\Delta P$ for the eruptions of 1999, 2010, 2016, and 2022.  This paper reports on 52 new eclipse times (for a total of 218 eclipses 1945--2025), plus a new theory result allowing for the confident calculation of $M_{\rm ejecta}$ from $\Delta P$.  The four eruptions ejected a total of (103$\pm$14)$\times$$10^{-6}$ $M_{\odot}$, while the white dwarf accreted 4$\times$$10^{-6}$ $M_{\odot}$ over the four previous eruption cycles.  With M$_{\rm ejecta}$=26$\times$M$_{\rm accreted}$, the U Sco white dwarf is losing large masses each eruption cycle, so U Sco can never produce a Type Ia supernova.

\end{abstract}

\section{INTRODUCTION}

Recurrent novae (RNe, Schaefer 2010) are a subset of classical novae (Chomiuk, Metzger, \& Shen 2020) that have a recurrence timescale ($\tau_{\rm rec}$) of under 100 years.  RNe are a class of cataclysmic variables (CVs) which are close interacting binaries, where a relatively ordinary companion star fills its Roche lobe and spills gas through an accretion disk onto a white dwarf (WD).  The accreted mass accumulates on the WD surface between nova eruptions (M$_{\rm accreted}$) until the trigger condition for thermonuclear runaway is reached, when the nova explosion ejects a shell of gas with mass M$_{\rm ejecta}$.  To have $\tau_{\rm rec}$$<$100 years\footnote{See equation 5 of Truran \& Livio 1986, and figure 7 of Shen \& Bildsten 2009.}, the CV must have a near-maximal white dwarf mass ($M_{\rm WD}$$>$1.2 $M_{\odot}$) and a near-maximal accretion rate ($\dot{M}$$\gtrsim$10$^{-7}$ $M_{\odot}$ yr$^{-1}$).  Only 11 RNe are known in our Milky Way, although there are undoubtedly hundreds more for which only zero-or-one nova eruptions have been discovered in the last century.

The famous prototypes of RNe are T CrB, T Pyx, and U Sco.  U Sco (Schaefer 2010, 2022a) was the first discovered with an eruption in 1863.  Further nova eruptions from U Sco where discovered in 1906, 1917, 1936, 1945, 1969, 1979, 1987, 1999, 2010, 2016, and 2022, for a total count of 12 eruptions.  With $\tau_{\rm rec}$$\sim$10 years, eruptions around 1926 and 1957 were likely missed due to the Sun passing through Scorpius and hiding any eruptions for several months each year.  For each eruption, U Sco rises in $\sim$4 hours from quiescence to a peak of $V$=7.5, fades by three magnitudes from peak in a time of $t_3$=2.6 days, and finally fades back to quiescent level with $V$=17.6, all within a total duration of 60 days.  U Sco displays {\it two} light curve plateaus in its otherwise smooth decline, for a light curve class\footnote{My definitions of light curve classes categorizes all the diversity of nova light curves into just seven classes (S, P, O, C, J, D, and F), as strongly correlated with the other observed properties and physics.  P-class nova display a plateau, or flattening, in the light curve around the time of transition, so U Sco is a P-class nova.  But U Sco, KT Eri, and V309 Del all have {\it two} plateaus, with this extra information optionally designated as class PP.} of PP(3).  The eruption spectra show it to be a He/N class, with the Balmer lines having a FWHM of 5700 km s$^{-1}$.  U Sco is an eclipsing binary with an orbital period $P$=1.23 days.  The WD must have $M_{\rm WD}$ close to 1.35 $M_{\odot}$.  The companion star has a size of 2.1 $R_{\odot}$, a surface temperature around 5000 K, and apparently a mass roughly at 1.0 $M_{\odot}$, being a subgiant star evolved past the main sequence.  After T CrB, U Sco is the best observed nova of any type, while the 2010 and 2022 eruptions are the two all-time best-observed nova events.

U Sco has long been the prototype for the most popular class of single-degenerate (SD) models for Type Ia supernova (SNIa) progenitor systems (e.g., Thoroughgood et al. 2001, Kato \& Hachisu 2012, Wang \& Han 2012).  The alluring logic is that RNe are WDs already near the Chandrasekhar mass with gas being piled on at a high rate, so the WD must soon get to the Chandrasekhar mass and explode as a SNIa.  The SNIa Progenitor Problem is one of the Grand Challenges that has high importance throughout many areas of astrophysics, and for which the astrophysics community has been sharply divided for the last four decades (Livio 2000, Maoz, Mannucci, \& Nelemans 2014, Ruiter \& Seitenzahl 2025).  Everyone agrees that the progenitor must be a close binary where one star is a carbon/oxygen (CO) WD that is somehow pushed to a mass where the carbon starts burning with enough energy to disrupt the exploding star.  Importantly, the WD cannot be an oxygen/neon (ONe) WD, as there is not enough energy available to blow up the star, so an accretion induced collapse will be the endpoint of the evolution.  Schematically, the solution can be characterized as being between two broad classes of models, the double-degenerate (DD) models and the SD models.  DD Models have the companion star being a second WD, hence the name `double-degenerate' for the two WDs in the binary.  These two WDs will in-spiral from gravitational radiation emission, and in the last second the two WDs merge or collide, providing the necessary total mass to initiate the nuclear burning.  SD models have the companion star being a relatively normal star, where the usual Roche lobe overflow drives the CO WD to the brink where it explodes.  There are a variety of versions for SD models, of which the popular models are for recurrent novae (with U Sco as the prototype), symbiotic stars (with T CrB as the prototype), and helium stars (with V445 Pup as the prototype).

The debate for whether U Sco is a progenitor is typically just the presentation of a theory model for evolution where the WD does or does-not increase in mass to approach the Chandrasekhar limit.  Unfortunately, this is all just theory calculations, where the community is sharply divided with contradictory claims, while there is scant observational evidence to support either side.  And theory for CV evolution can make RNe either with or without CO WDs.  Further, for secondary indications (like RNe formation rates, galactic cosmochemistry calculations) theory models give contradictory claims.  So no confident solution is in sight from theory and models.  What is needed are observational tests of the primary properties.  There are two such tests, first whether U Sco and RNe have their $M_{\rm WD}$ increasing-or-decreasing over time, and whether their WDs are CO or ONe composition.

The basic situation was obvious back in 1988, when one of us (Schaefer) started a career-long program to measure the $M_{\rm ejecta}$ for U Sco, then for many of the galactic RNe, then generalized to include many classical novae.  The idea was to measure the orbital period {\it before} the eruption ($P_{\rm before}$) and then measure the orbital period {\it after} the eruption ($P_{\rm after}$), where the period change across the eruption ($\Delta P$=$P_{\rm after}$-$P_{\rm before}$) is simply related to $M_{\rm ejecta}$ by Kepler's Law.  This program has the strong advantage that it is a simple timing experiment capable of good accuracy, with no problems from the usual pernicious uncertainties (like from reddening, distance, filling factors,...).  This program has the strong disadvantage that it takes decades of persistent observing runs from year-by-year to measure $\Delta P$, and the program can only be run for novae that are moderately bright in quiescence and with some phase marker (like eclipses or prominent ellipsoidal effects).  For patient observers willing to devote large amounts of telescope time year-by-year, the result is a reliable and accurate measure of $M_{\rm ejecta}$.  This is to be combined with independent measures of the mass accreted between eruptions, with these coming from several independent methods with enough accuracy to answer the question.  Then, the answer comes from whether $M_{\rm ejecta}$ is larger-or-smaller than $M_{\rm accreted}$.  If $M_{\rm ejecta}$ is larger, then the WD must be losing mass over each eruption cycle, so the star cannot be a SNIa progenitor.

To start this program, Schaefer (1988, 1990) discovered the eclipses of U Sco, and then discovered the orbital periods of RNe T Pyx and V394 CrA (Schaefer 1990, 2009, Schaefer et al. 1992).  U Sco eclipse times have been reported in Schaefer (1990, 2011, 2022a), Schaefer \& Ringwald (1995), and Schaefer et al. (2011).  This paper reports on 52 new eclipse times, bringing us up to 2025.4.  In the meantime, for T CrB, recovery of 213,730 $B$ and $V$ magnitude from 1842 to 2022 allowed for the derivation of $\Delta P$ across the 1946 eruption (Schaefer 2023a), while an exhaustive analysis of all radial velocity measures 1947--2025 allowed for small improvements in the T CrB $\Delta P$ (Schaefer 2025c).  For T Pyx, a highly accurate $\Delta P$ has been measured across its 2011 eruption (Schaefer et al. 2013, 2023b).  For CI Aql, a poor measure of $\Delta P$ has been derived for its 1999 eruption (Schaefer 2011, 2023b), with the problem being that scant pre-eruption data have been found.  For the unique helium nova V445 Pup, Schaefer (2025a) measured an accurate period change across its 2000 eruption.  For six other classical novae (QZ Aur, HR Del, DQ Her, BT Mon, RR Pic, and V1017 Sgr), $\Delta P$ values have been measured, as reported in Schaefer (2020a, 2020b; 2024), Schaefer \& Patterson (1983), Schaefer et al. (2019), and Salazar et al. (2017).  With all this work, we have good experience at pulling out reliable measures of $\Delta P$.

In this paper, we concentrate on the new results for U Sco.  We now have 218 eclipse times from 1945 to 2025.4, with the result that we have good measures of the $\Delta P$ for the eruptions of 1999.2, 2010.1, 2016.78, and 2024.4.  A critical advance in this paper is the realization that the various mechanisms for angular momentum loss across the eruption are all of negligible size, so we can translate accurately and reliably from $\Delta P$ to $M_{\rm ejecta}$.  In the end, we measure confidently that the ejecta mass is 26$\times$ the accreted mass, so the U Sco WD must be losing mass over each eruption cycle.

\section{NEW LIGHT CURVE DATA}

\subsection{Siding Spring Myers Observatory}

\begin{figure}
\epsscale{1.01}
\plotone{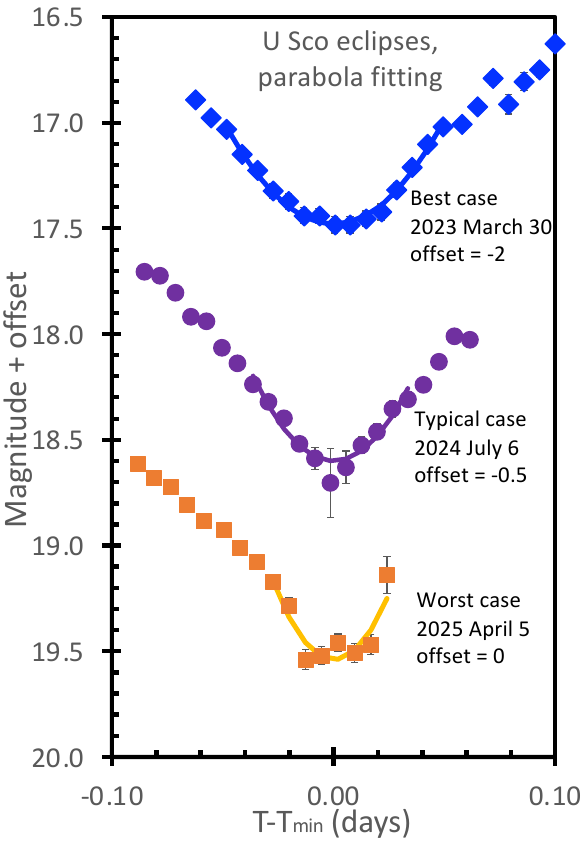}
\caption{Three U Sco eclipses.  For each eclipse, our measured light curves are in the $CV$ band, with most photometric error bars being smaller than the plotting symbol.  For each eclipse, we have fit a parabola by the usual chi-square minimization calculation, with the best-fitting parabola shown superposed on each light curve.  To illustrate our eclipse light curves, we show our best case (top curve), typical case (middle curve), and worst case (bottom curve).}
\end{figure}

The most important of the new U Sco data are our many time series covering eclipses from 2022 to 2025.  These are all taken by one of the authors (Myers), with remote observing telescope time from Australia.  This remote observing procedure is what allows for measuring the large number of eclipses, whereas the older style involves long travel and U Sco eclipses once every five nights (if clear) from Cerro Tololo.  Importantly, this remote observing allows for catching many eclipses over many nights spread out over several months for every year, with this being critical for measuring period changes of U Sco.  Critically, our 45 eclipse times are largely the only coverage for the seven observing seasons 2019--2025.  These provide the only information on the sudden period change across the 2022 eruption ($\Delta P$) and the steady period change after the 2022 eruption ($\dot{P}$), with these measures being a primary goal of this paper.

The main telescope is a Planewave CDK 0.41-m located in Coonabarabran Australia.  The CCD is an FLI PL4710.  We always observe U Sco light curves with an Astrodon Clear filter, so as to maximize the signal.  Our large compilations of U Sco eclipse times in a wide variety of bands proves that the $O-C$ values have no color dependency, so running filterless is best.  The calibration of the magnitudes is versus comparison stars with V-band magnitudes, with this being labelled `$CV'$.  The typical photometric uncertainty is $\pm$0.02 mag outside of eclipse, and $\pm$0.05 mag at the times of minima.  The time series are run remotely with automated scripts.  The exposure times are typically 10 minutes, which is adequate to fully-resolve the U Sco eclipse.  The typical run duration is 4-6 hours, centered on mid-eclipse.

All of our Coonabarabran time series data for U Sco are publicly available from the archive of the American Association of Variable Star Observers (AAVSO)\footnote{\url{https://www.aavso.org/data-download}}.  The AAVSO observer ID for Myers is MGW.  Figure 1 shows our light curves for three eclipses.  To show the range of quality, we have plotted our best, our typical, and our worst eclipse light curves.  The quality of the case depends primarily on the time range of coverage, which usually is governed by clouds and the rise and set times of U Sco from our observatory.

\subsection{$AAVSO$}

Nearly all Galactic nova light curves have the large majority of data taken by small-telescope observers from around the world (Strope, Schaefer, \& Henden 2010).  This vast data collection program is orchestrated and archived by the AAVSO.  Since the year 2000 or so, the bulk of the data have been with CCDs, usually through standard $UBVRI$ filters, and with professional quality acquisition, comparison stars, and reduction.  

Up until the year 2010, nova light curve observations only consisted of many individual observers taking one or two observations per night, night-after-night for the entire duration of the eruption (and beyond).  Such is required to outline the shape and properties of the light curve, as required for many science questions.  But no one made long photometric time series of novae {\it during} eruptions.  This is somewhat surprising because both amateur and professional observers had long been making long time series photometry of novae in quiescence.  Perhaps the reason is that no one had anticipated that novae in eruption could be variable on fast time scales.  With a large scale program prepared {\it in advance} specifically for the 2010 eruption of U Sco, the AAVSO orchestrated a massive campaign involving hundreds of photometric observers around the world (Schaefer et al. 2010).  This resulted in a light curve with average coverage of {\it once every 2.6 minutes throughout the entire eruptions from peak to quiescence} (Schaefer et al. 2011, Pagnotta et al. 2015).  Suddenly, with the first time series of a nova eruption, two new and unexpected phenomena (fast flares around transition and late time eclipses from a puffed up disk) were discovered.

For the 2022 U Sco eruption, the AAVSO observers repeated the massive 2010 observing program.  The AAVSO archive contains 42,295 magnitudes over the entire 60-days of the U Sco eruption, for an average of one measure every 2.0 minutes.  This is the database that we use in this paper for deriving eclipse times and light curve properties during the 2022 eruption, as well as the average $V$ magnitude in quiescence after 2022.

\subsection{$TESS$}

\begin{figure*}
\epsscale{1.17}
\plotone{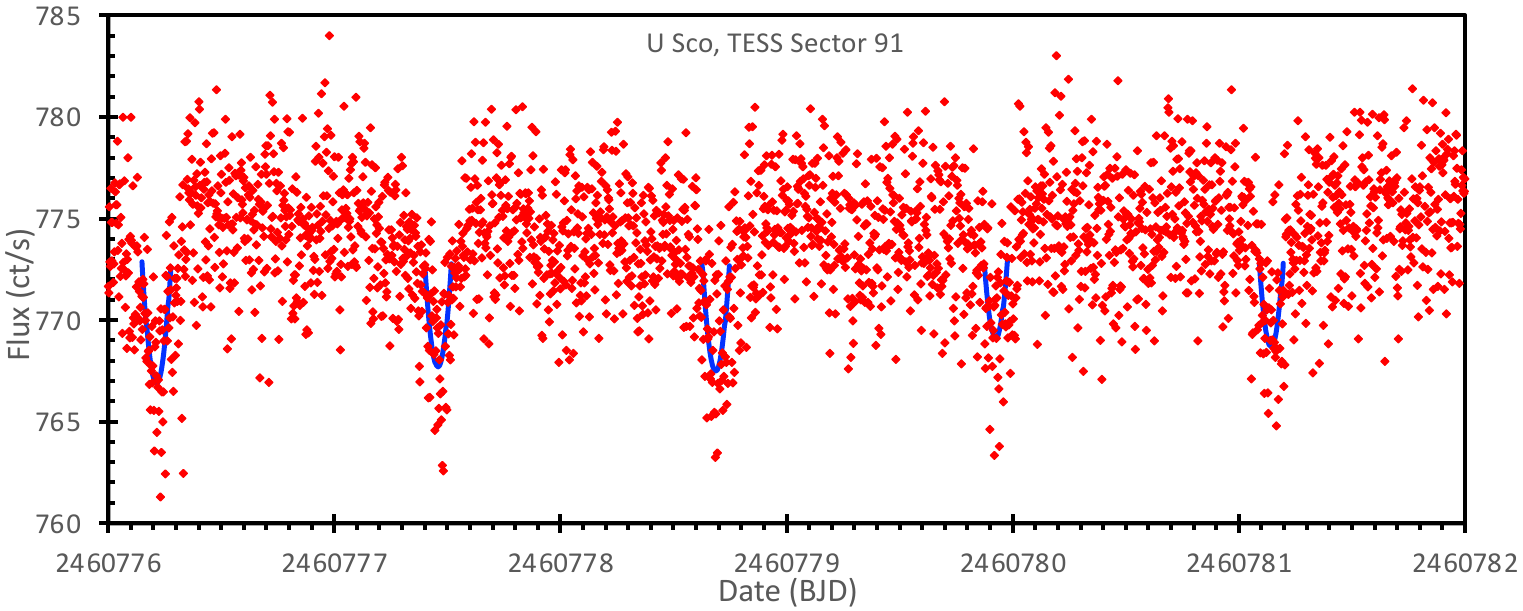}
\caption{TESS light curve for U Sco.  This light curve is from the first orbit of Sector 91 of {\it TESS} in 2025 April, with 200 second time bins.  The Poisson measurement error is $\pm$2.6 ct/s for each data point, which is comparable to the ordinary flickering noise always present in the U Sco light curve.  The constant background flux level is unknown, because the large pixel size (21''$\times$21'') contains other stars and a lot of background light, and this makes the amplitude of eclipse to appear to be small.  This first orbit shows 5 eclipses, with the best fit parabolas shown as blue curves.  The TESS eclipse light curves are poor because U Sco is near the limit of detection and because the background levels are high due to the pixel size.  The second orbit of Sector 91 (i.e., the second half of the sector) has 6 eclipses, with these being substantially poorer in quality.}
\end{figure*}

The {\it TESS} satellite is producing awesome nearly-gap-free light curves lasting just under one month in duration for most stars in the sky down to around 18th mag (Ricker at al. 2015).  The sky coverage is in `Sectors', each consisting of roughly 27 days of nearly-gap-free time series, each Sector lasting for two orbits of the spacecraft.  As this is being written, $TESS$ has just completed 91 Sectors of observations, with most stars being covered with 2-6 Sectors of photometry.  For nova studies, this provides a wonderful and unique resource for the discovery of new orbital periods (Schaefer 2022b) and for measuring the steady period changes, $\dot{P}$ (Schaefer 2023b).  A feature of the $TESS$ sky coverage is that the Galactic center region was not covered at all during the first 90 Sectors, only to have the 91st Sector covering the entire center.  This means that U Sco has received no $TESS$ coverage up until the month of April 2025, and it will receive no coverage after this Sector 91.  So we have been anxiously awaiting the downloading of Sector 91 data.

The $TESS$ data are downloaded and processed by the Barbara A. Mikulski Archive for Space Telescopes (MAST), with free public access\footnote{\url{https://mast.stsci.edu/portal/Mashup/Clients/Mast/Portal.html}}.  The U Sco data have a time resolution of 200 seconds.  The valid data consist of 7 days starting 2025 April 9, plus 7 days starting 2025 April 23.  This is 5872 measured fluxes, in units of counts per second.  To extract the light curve, we use the standard program LightKurve.  After experimentation, we get the best signal with a photometry aperture that is only one pixel in size, with this being 21$\times$21 arc-seconds.  We used the PLD algorithm, yet still had to empirically remove large instrumental trends.  The resultant light curve for the first orbit of Sector 91 is shown in Figure 2.

The figure shows five eclipses.  The noise is huge, being somewhat larger than the eclipse amplitude.  The reason for this is that U Sco is faint, down near the detection limit.  Further, the single large pixel contains background stars and substantial skylight, so the normal Poisson noise is much larger than desired.  Still, despite this poor showing, the $TESS$ light curve does provide 11 eclipse times.

\section{ERUPTION LIGHT CURVE}

\subsection{Mysterious Flares Just Before the Transition}

\begin{figure*}
\epsscale{1.17}
\plotone{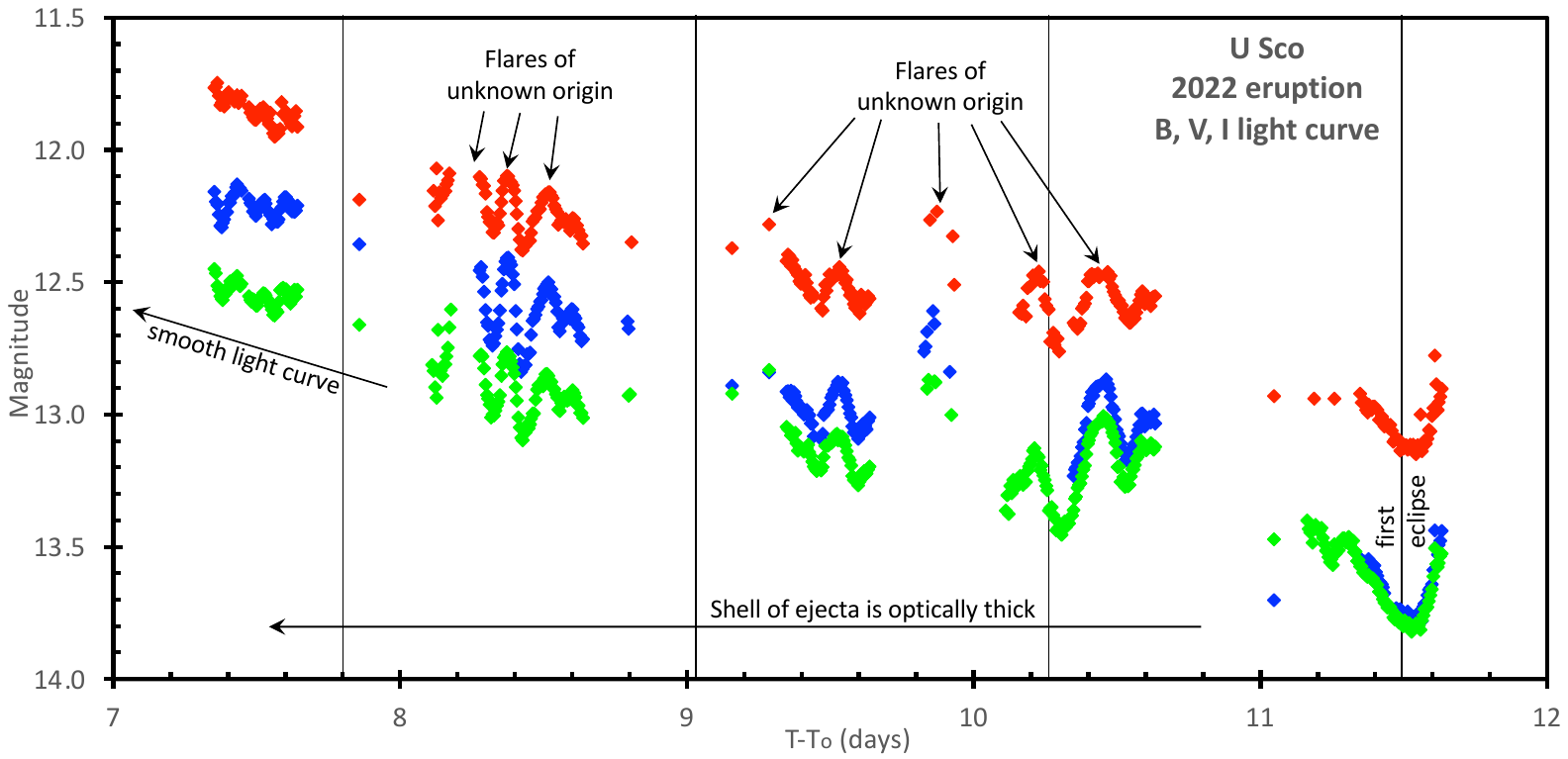}
\caption{Mysterious flares just before transition.  This light curve shows only $B$, $V$, and $I$ (blue, green, and red diamonds respectively) for Days $+$7 to $+$12 after the start of the eruption, $T_0$, taken to be HJD 2459737.2.  The observers contributing to these light curves are F.-J. Hambsch (with AAVSO observer ID of HMB), B. Harris (HBB), S. Dvorak (DKS), B. Monard (MLF), W. Cooney (CWT), M. Richmond (RHM), A. Pearce (PEX), R. Schmidt (SREB), S. M. Brincat (BSM), M. Larsson (LMN), C. Galdies (GCHB), and G. Myers (MGW).  The times of zero orbital phase are represented by the thin vertical lines spaced every 1.23 days.  The first eclipse seen during the eruption is centered on Day $+$11.5, while no eclipse is seen centered one orbit earlier.  So the shell of ejecta is optically thick before Day $+$11.  Before Day $+$8, the fading light curve is smooth.  But from Days $+$8 to $+$11, U Sco suffered at least 8 flares with amplitudes 0.2--0.5 mags.  These flares are fast, with rise times of 0.035--0.090 days, and durations of 0.110--0.165 days.  Critically, the extra flare light is extremely blue, with a dereddened $B-V$ from $-$0.6 to $-$0.8.  This can be seen by noting that the flare amplitude in $B$ is much larger than in $V$ and $R$.  The extremity of this color rules out the various tries at explanation.   }
\end{figure*}

The transition phase of a nova eruption is when the shell of ejecta becomes optically thin due to the dilution of the gas from expansion.  This is the time that the light curve breaks from a fast decline to a slow decline, making a distinct flattening of the light curves.  The transition marks the time when the spectrum changes from a photosphere with emission lines to an emission line dominated Nebular spectrum.  The transition time is when the central binary star becomes visible through the thinning murk of the expanding shell, and this is when eclipses can become visible.  

During the 2010 U Sco eruption, the transition was observed to be within a day or two of $+$13.68 days after the start of the eruption (Schaefer et al. 2011).  For all the time from the peak up until near the transition, the light curve was smooth, to within the measurement errors.  Startlingly, from Day $+$8.6 to $+$12, U Sco had flares with amplitudes 0.5 mag above the trend line and with rise and fall times of 0.6--1.2 hours.  No one had ever seen such a phenomenon, largely because no on previously had any substantial time series on a nova in eruption.

In 2022, the U Sco light curve break appeared 12--15 days after the start, while the first shallow eclipse was seen on Day 11.5.  Figure 3 shows a blow-up of the light curve with only the $B$, $V$, and $I$ data shown.  Before Day $+$8, the light curve looks smooth, as expected.  However, from Days $+$8.2 to $+$10.4, 6 flares are seen with amplitudes 0.2--0.5 mag, rise times 0.035--0.090 days, and durations 0.110--0.165 days.  These flares do not display any significant periodicity.  These flares are highly significant, reported simultaneously and identically by time series from multiple observers, and measured with the same variations independently in $B$, $V$, and $I$.  The observers of these flaring time series are amongst the best and most-experienced nova photometrists in the world.  There is no doubt about the existence and properties of these flares.

The transition-flaring seen in 2022 confirms their discovery in 2010, and provides further measures.  These transition flares are sharply constrained to Days $+$8.2 to $+$10.7 for the 2022 eruption and Days $+$8.7 to $+$11.7 for the 2010 eruption.  This restricted time range should be a strong clue and requirement for theory models.  

We can also measure the $B-V$ colors.  The $B-V$ for the light curve is near 0.0 up to Day 7, as is usual for nova eruptions.  Then the non-flaring color changes to -0.3 at the start of Figure 3, for unknown reasons that can only be due to especially-hot gas appearing at the photosphere.  For the flare light alone, we can tabulate the magnitudes at the start and peak of each flare.  The excess flare light has $B-V$ from $-$0.4 to $-$0.6 mag, which is dereddened to $-$0.6 to $-$0.8 mag.  The excess light has $V-I$ from $+$0.3 to $+$0.5.  These colors are greatly different from any power-law or blackbody.  The flares have an incredibly strong blue-excess.  This points to the flares as being from a mechanism that is extremely `hot'.

There actually is precedent for novae having flares before the transition, and that is the many jitters in the J-class novae (Strope, Schaefer, \& Henden 2010).  Connecting to jitters would not be helpful, because their mechanism is not known.  Nevertheless, the J-class jitters are greatly different from the U Sco transition-flares in all the critical measures.  Individual jitters have durations 3--30 days (rather than $<$0.165 days), whilst the duration of the phase with jitters is 20--160 days (rather than 2 days).  The jitters only flare before-peak and around-peak (rather than $\sim$6 mags below the peak).  The jitter light has extinction-corrected $B-V$ from 0.0 to $+$0.4 mag, being similar in color to the base photosphere (rather than $-$0.6 to $-$0.8 mag).  Jitters are only known to occur on the lowest mass white dwarfs (0.6--0.8 $M_{\odot}$), while U Sco is near the highest possible mass.  It is not plausible that jitters and transition-flares have the same mechanism.  

What can be the cause of these flares?  The flares are during times when the shell is optically thick, so the flares must be caused by some mechanism associated with the expanding gas shell.  Our first speculation was that random transverse motion of the gas cloudlets might create  brief openings to reveal the inner binary.  But this is refuted by the extreme color of the flare light (with $B-V$ around $-$0.7), while we know from a few days later that the inner binary is not so blue in color.  Our second speculation was that the inner binary sent out some number of high velocity shells that collide with each other, much like the internal shocks in Gamma-Ray Bursts, to create flares.  But this idea fails to allow for the extreme blue colors, being bluer than ever seen in any nova, supernova, or Gamma-Ray Burst.  Further, the fast decline in this scenario could only be from recombination, but that time scale is greatly longer than 0.05 days or so, so any shocked gas cannot fade as fast as observed.  Our third speculation was that the variability is from two adjacent dips making a flare appear between.  But this idea fails because the light between the dips still has the extreme color $B-V$ of $-$0.7 mag which is not seen at any other time.  Our fourth speculation is that the inner binary is modulating the density of the ejecta, so that a flare appears when some particularly dense shell of ejecta passes through the photosphere.  But this idea fails due to the extreme $B-V$ color, as such cannot be produced simply by increasing the density of some shell within the ejecta.  So we have rejected all the ordinary scenarios to explain the transition-flares, and hence we label them as `mysterious'.  But U Sco has solved this problem with some now-unknown mechanism, so explaining these mystery-flares is now a challenge for theorists.

\subsection{2010 Versus 2022 Light Curve Comparison}

\begin{figure}
\epsscale{1.17}
\plotone{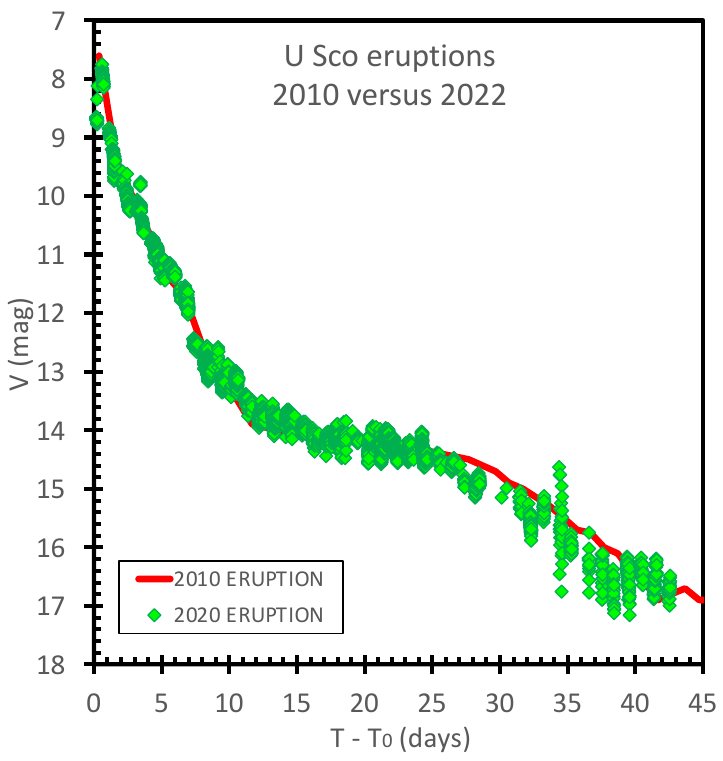}
\caption{2010 \& 2022 light curves.  An important question is whether all eruptions of U Sco have identical light curves.  Schaefer (2010) made an exhaustive comparison of 7 U Sco events before 2000 that are identical to within the measurement errors.  Now, the 2010 and 2022 light curves are the two all-time best nova light curves (covering the entire eruption, with average frequency of under 2.6 minutes per magnitude throughout, with much $BVRI$ measures), so a detailed comparison will show even small differences.  (Here, we do not consider differences due to ordinary flickering, due to the exact phasing of eclipses, nor due to the quiescent level being asymptotically approached.)  Here, we plot the light curves for the two eruption, with the 2010 light curve taken from the $V$-band out-of-eclipse trend line from Schaefer (2011), and the green diamonds for all the AAVSO $V$ magnitudes with orbital phases from 0.1--0.9.  For most of the duration of the eruption, the red curve is completely hidden under the green diamonds.  At later times, the two curves asymptote to slightly different quiescent levels, so that the two light curves are identical when we only consider the eruption light.  The small differences between the 2010 and 2022 light curves are consistent with the ordinary problems of comparing observations from groups of observers with differing comparison stars and color terms.  That is, the 2010 and 2022 eruptions have identical light curves, with no measurable differences.  This creates a conundrum because all the eruptions are photometrically (and spectroscopically) identical, yet the eruptions result in greatly different $\Delta P$ and $\dot{P}$ behavior.  So the mechanism that changes the $P$ must be well hidden in the binary.   }
\end{figure}

Schematically, all eruptions from an individual RN should be identical.  (This is because the eruptions depend primarily on the white dwarf mass, the accretion rate, and the composition of the companion star, where these are constant from eruption-to-eruption.)  We have found that the U Sco period changes ($\Delta P$ and $\dot{P}$) vary from eruption-to-eruption, so there must be {\it some} major difference between eruptions.  Whatever this difference is, we expect that this might be manifest as differences in the eruption light curves.  So to understand the period changes, it would be good to measure the differences in the light curves.

Schaefer (2010) made an exhaustive analysis of {\it all} light curves for all ten Galactic RNe, finding that they were all identical to within measurement errors.  For U Sco specifically, seven eruption light curves were compared and found to be identical, to within the normal measurement errors.

The 2010 eruption has the all-time-best nova eruption light curve, and now the AAVSO light curve for the 2022 eruption is equally good.  This allows for a nice comparison between two excellent light curves.  Muraoka et al. (2024) started this comparison.  Unfortunately, their light curves have a number of the usual small troubles.  In particular, they superposed offset magnitudes from many bands and this leads to systematic errors, they included magnitudes from eclipse times, and they made no heliocentric corrections for eclipse times (dominated by magnitudes near the out-of-eclipse level).  To make a better comparison, in Figure 4, we have over-plotted the $V$-band trend line (constructed with non-eclipsing phases and correct $V$-band measures) with the $V$-band observations (constructed with non-eclipsing phases and only the $V$-band measures).  These two light curves should be directly comparable.  What we see is that the 2022 light curve overlaps the red 2010 light curve to the extent that much of the red light curve is hidden.  In particular, the plateau in 2010 started on the same Day as in 2022.  That is, to within the intrinsic variability (flickering) and measurement errors, the 2010 and 2022 light curves are identical.

Evans et al. (2023) makes a detail comparison of the $H$-band near-infrared light curves from 2010 and 2022.  These are found to be identical.

The 2010 and 2022 eruptions can also be compared by their X-ray properties.  The {\it Swift} X-ray light curves for both 2010 and 2022 were constructed by K. Page (University of Leicester), and plotted in Muraoka et al. (2025).  The X-ray light curves and hardness ratio curves are largely the same.  But between Days $+$12 to $+$15, the 2022 X-ray flux was fainter than in 2010 by a factor of  $\sim$50.  Muraoka interprets this as the soft X-ray stage beginning around three days earlier in 2010 than in 2022.  This start phase is related to the thinning of the ejecta shell allowing the X-rays of the inner binary to shine through.

A third way to time the transition is to record the Day of the first eclipse.  In 2010, the first eclipse was seen on Day $+$13.68 (Schaefer et al. 2010), although it is possible that earlier eclipses were not covered.  In 2022, the first eclipse was certainly on Day $+$11.5, and certainly not visible from one orbit earlier.

Now we have three measures of the transition time, with the 2010 transition at the same time (from the start of the plateau in both the optical and near-infrared), three days earlier (from the X-ray phasing), and two days later (from the first eclipses).  These differences appear to be from intrinsic variations and errors.  With the three measures, we conclude that the transition times were not significantly different between the 2010 and 2022 eruptions.

So all U Sco eruptions have identical light curves (to within measurement errors).  This then emphasizes the problem of explaining what mechanism can possibly make for large eruption-to-eruption $\Delta P$ and $\dot{P}$ changes.  How can the period-changes have big differences while their light curves have no measurable differences?

\section{ECLIPSE TIMES}

The U Sco eclipses can be measured from time series photometry, with the goal of measuring the instant of the inferior conjunction.  In principle, for a symmetrical accretion disk with the center being brightest, the time of eclipse minimum is the time of conjunction.  There are a variety of methods to derive the time of minimum.  Schaefer (2021) made a detailed study of 166 U Sco eclipse times, plus  1567 time of minima for DQ Her, RW Tri, and UX UMa, with these quantifying the RMS variations between all pairs of eclipse times.  ``The worst measures by far are those involving the brighter half of the eclipse, and the time bin of minimum light.  The best measures (being roughly equal in quality) are for the fitted parabolas to the minimum light curve and the bisectors cutting across the eclipse profile at roughly 50\% to 80\% of the depth of minimum.''  This is the justification for our consistent use of parabola fitting to measuring eclipse times.  Of comparable importance is that we have used the same methodology for all eclipse times, so that there will be no systematic offsets produced artificially in the $O-C$ diagram.

Examples of three eclipsing light curves and the parabolic fits are shown in Figure 1.  We know from the excellent averaged light curve from {\it K2} that the real shape of the U Sco eclipse minimum is exactly a parabola (Schaefer 2022a, figure 5).  The deviations from a parabola are due to the ordinary flickering.  (This flickering is still prominent throughout the minimum.)  This flickering will skew the eclipse times early or late.  The eclipse profiles are often asymmetric for times around the start of the ingress and the end of the egress, likely due to large-scale structure in the accretion disk.  These broad asymmetries vary from eclipse-to-eclipse, with this adding large errors to eclipse times derived using these times\footnote{The times in Muraoka et al. (2024) are not reliable, even if they are made to include the heliocentric correction.}.  This is just further confirmation that our parabolic fits to find the times of minima is the best ways to get the times of conjunction.

\startlongtable
\begin{deluxetable}{lllrr}
\tablecaption{218 eclipse times for U Sco 1945--2025}
\tablewidth{600pt}
\tabletypesize{\scriptsize}
\tablehead{
\colhead{Year}   &   \colhead{Telescope}  &  \colhead{$T$$\pm$$\sigma_{\rm meas}$ (HJD)}   &   \colhead{$N$}   &   \colhead{$O$-$C$ }         
}
\startdata
1945.503	&	Harvard	&	2431639.3000	$\pm$	0.0090	&	-15924	&	-0.0091	\\
1989.523	&	CT 0.9m	&	2447717.6064	$\pm$	0.0062	&	-2858	&	-0.0291	\\
1989.526	&	CT 0.9m	&	2447718.8481	$\pm$	0.0084	&	-2857	&	-0.0180	\\
1989.536	&	CT 0.9m	&	2447722.5406	$\pm$	0.0018	&	-2854	&	-0.0171	\\
1989.539	&	CT 0.9m	&	2447723.7675	$\pm$	0.0030	&	-2853	&	-0.0208	\\
1989.550	&	CT 0.9m	&	2447727.4707	$\pm$	0.0052	&	-2850	&	-0.0092	\\
1990.415	&	Wade	&	2448043.7262	$\pm$	0.0042	&	-2593	&	-0.0043	\\
1993.458	&	KP 0.9m	&	2449154.9116	$\pm$	0.0026	&	-1690	&	-0.0028	\\
1994.563	&	CT 0.9m	&	2449558.5258	$\pm$	0.0020	&	-1362	&	-0.0080	\\
1994.566	&	CT 0.9m	&	2449559.7605	$\pm$	0.0046	&	-1361	&	-0.0038	\\
1994.576	&	CT 0.9m	&	2449563.4515	$\pm$	0.0069	&	-1358	&	-0.0045	\\
1995.482	&	CT 0.9m	&	2449894.4733	$\pm$	0.0023	&	-1089	&	0.0002	\\
1995.486	&	CT 0.9m	&	2449895.6939	$\pm$	0.0020	&	-1088	&	-0.0097	\\
1995.499	&	CT 0.9m	&	2449900.6196	$\pm$	0.0011	&	-1084	&	-0.0062	\\
1996.564	&	CT 0.9m	&	2450289.4682	$\pm$	0.0022	&	-768	&	-0.0104	\\
1997.356	&	CT 0.9m	&	2450578.6517	$\pm$	0.0012	&	-533	&	-0.0055	\\
1997.359	&	CT 0.9m	&	2450579.8951	$\pm$	0.0034	&	-532	&	0.0074	\\
1997.369	&	CT 0.9m	&	2450583.5677	$\pm$	0.0035	&	-529	&	-0.0117	\\
1999.205	&	Matsumoto	&	2451254.2110	$\pm$	0.0100	&	16	&	-0.0165	\\
1999.236	&	VSNET	&	2451265.3060	$\pm$	0.0100	&	25	&	0.0036	\\
1999.293	&	AAT	&	2451286.2143	$\pm$	0.0050	&	42	&	-0.0074	\\
2001.335	&	McD 2.1m	&	2452031.9339	$\pm$	0.0014	&	648	&	0.0008	\\
2001.594	&	McD 2.1m	&	2452126.6874	$\pm$	0.0010	&	725	&	0.0022	\\
2001.607	&	McD 2.1m	&	2452131.5975	$\pm$	0.0090	&	729	&	-0.0099	\\
2002.413	&	McD 2.1m	&	2452425.7093	$\pm$	0.0029	&	968	&	0.0012	\\
2002.426	&	McD 2.1m	&	2452430.6296	$\pm$	0.0030	&	972	&	-0.0008	\\
2002.429	&	McD 2.1m	&	2452431.8552	$\pm$	0.0041	&	973	&	-0.0057	\\
2003.349	&	McD 2.7m	&	2452767.7993	$\pm$	0.0007	&	1246	&	-0.0009	\\
2003.406	&	McD 0.8m	&	2452788.7162	$\pm$	0.0011	&	1263	&	-0.0033	\\
2003.423	&	MDM 1.3m	&	2452794.8720	$\pm$	0.0008	&	1268	&	-0.0002	\\
2003.481	&	MDM 1.3m	&	2452815.7849	$\pm$	0.0008	&	1285	&	-0.0066	\\
2003.508	&	McD 2.7m	&	2452825.6420	$\pm$	0.0029	&	1293	&	0.0061	\\
2004.501	&	CT 1.0m	&	2453188.6481	$\pm$	0.0019	&	1588	&	0.0008	\\
2004.515	&	CT 1.0m	&	2453193.5675	$\pm$	0.0011	&	1592	&	-0.0019	\\
2005.438	&	CT 0.9m	&	2453530.7350	$\pm$	0.0030	&	1866	&	-0.0043	\\
2005.465	&	CT 0.9m	&	2453540.5830	$\pm$	0.0038	&	1874	&	-0.0007	\\
2006.344	&	CT 0.9m	&	2453861.7560	$\pm$	0.0009	&	2135	&	-0.0004	\\
2006.432	&	CT 0.9m	&	2453893.7513	$\pm$	0.0019	&	2161	&	0.0006	\\
2006.503	&	CT 0.9m	&	2453919.5932	$\pm$	0.0018	&	2182	&	0.0011	\\
2006.691	&	CT 0.9m	&	2453988.5038	$\pm$	0.0009	&	2238	&	0.0010	\\
2007.382	&	CT 0.9m	&	2454240.7660	$\pm$	0.0008	&	2443	&	0.0011	\\
2007.685	&	CT 0.9m	&	2454351.5159	$\pm$	0.0005	&	2533	&	0.0018	\\
2008.346	&	CT 0.9m	&	2454592.7013	$\pm$	0.0008	&	2729	&	0.0000	\\
2008.376	&	CT 0.9m	&	2454603.7787	$\pm$	0.0015	&	2738	&	0.0025	\\
2008.433	&	CT 0.9m	&	2454624.6950	$\pm$	0.0011	&	2755	&	-0.0005	\\
2008.490	&	CT 0.9m	&	2454645.6193	$\pm$	0.0005	&	2772	&	0.0045	\\
2009.238	&	CT 0.9m	&	2454918.7959	$\pm$	0.0005	&	2994	&	-0.0004	\\
2009.296	&	CT 0.9m	&	2454939.7113	$\pm$	0.0009	&	3011	&	-0.0043	\\
2009.339	&	CT 0.9m	&	2454955.7163	$\pm$	0.0010	&	3024	&	0.0036	\\
2009.484	&	CT 0.9m	&	2455008.6305	$\pm$	0.0048	&	3067	&	0.0043	\\
2009.528	&	McD 82"	&	2455024.6217	$\pm$	0.0027	&	3080	&	-0.0016	\\
2009.572	&	CT 0.9m	&	2455040.6145	$\pm$	0.0007	&	3093	&	-0.0059	\\
2010.118	&	Stein	&	2455239.9600	$\pm$	0.0200	&	3255	&	-0.0090	\\
2010.131	&	Oksanen	&	2455244.8778	$\pm$	0.0005	&	3259	&	-0.0134	\\
2010.138	&	Tan	&	2455247.3505	$\pm$	0.0018	&	3261	&	-0.0018	\\
2010.145	&	Oksanen	&	2455249.8047	$\pm$	0.0008	&	3263	&	-0.0087	\\
2010.151	&	Stockdale	&	2455252.2681	$\pm$	0.0013	&	3265	&	-0.0064	\\
2010.175	&	Oksanen	&	2455260.8838	$\pm$	0.0010	&	3272	&	-0.0045	\\
2010.188	&	Oksanen	&	2455265.8097	$\pm$	0.0015	&	3276	&	-0.0008	\\
2010.195	&	Stockdale	&	2455268.2625	$\pm$	0.0020	&	3278	&	-0.0091	\\
2010.202	&	Oksanen	&	2455270.7446	$\pm$	0.0009	&	3280	&	0.0119	\\
2010.205	&	Krajci	&	2455271.9637	$\pm$	0.0031	&	3281	&	0.0005	\\
2010.232	&	Oksanen	&	2455281.8158	$\pm$	0.0012	&	3289	&	0.0082	\\
2010.246	&	Oksanen	&	2455286.7411	$\pm$	0.0025	&	3293	&	0.0113	\\
2010.377	&	CT 0.9m	&	2455334.7211	$\pm$	0.0009	&	3332	&	0.0000	\\
2010.492	&	CT 0.9m	&	2455376.5650	$\pm$	0.0035	&	3366	&	0.0053	\\
2010.508	&	MDM 2.4m	&	2455382.7126	$\pm$	0.0008	&	3371	&	0.0001	\\
2010.522	&	CT 0.9m	&	2455387.6395	$\pm$	0.0010	&	3375	&	0.0048	\\
2010.623	&	Oksanen	&	2455424.5565	$\pm$	0.0010	&	3405	&	0.0054	\\
2011.283	&	CT 0.9m	&	2455665.7492	$\pm$	0.0009	&	3601	&	0.0109	\\
2011.341	&	CT 0.9m	&	2455686.6672	$\pm$	0.0016	&	3618	&	0.0096	\\
2011.401	&	CT 0.9m	&	2455708.8169	$\pm$	0.0015	&	3636	&	0.0095	\\
2011.442	&	CT 0.9m	&	2455723.5826	$\pm$	0.0010	&	3648	&	0.0086	\\
2011.485	&	CT 0.9m	&	2455739.5762	$\pm$	0.0012	&	3661	&	0.0051	\\
2011.560	&	CT 0.9m	&	2455766.6569	$\pm$	0.0009	&	3683	&	0.0138	\\
2012.233	&	CT 0.9m	&	2456012.7698	$\pm$	0.0007	&	3883	&	0.0173	\\
2012.335	&	CT 0.9m	&	2456049.6833	$\pm$	0.0010	&	3913	&	0.0144	\\
2012.409	&	CT 0.9m	&	2456076.7509	$\pm$	0.0008	&	3935	&	0.0100	\\
2012.567	&	CT 0.9m	&	2456134.5890	$\pm$	0.0009	&	3982	&	0.0123	\\
2012.611	&	CT 0.9m	&	2456150.5850	$\pm$	0.0010	&	3995	&	0.0112	\\
2014.649	&	K2	&	2456895.0641	$\pm$	0.0010	&	4600	&	0.0094	\\
2014.652	&	K2	&	2456896.2991	$\pm$	0.0009	&	4601	&	0.0139	\\
2014.659	&	K2	&	2456898.7582	$\pm$	0.0009	&	4603	&	0.0119	\\
2014.663	&	K2	&	2456899.9905	$\pm$	0.0007	&	4604	&	0.0136	\\
2014.666	&	K2	&	2456901.2212	$\pm$	0.0007	&	4605	&	0.0138	\\
2014.669	&	K2	&	2456902.4519	$\pm$	0.0009	&	4606	&	0.0139	\\
2014.673	&	K2	&	2456903.6797	$\pm$	0.0009	&	4607	&	0.0112	\\
2014.676	&	K2	&	2456904.9121	$\pm$	0.0009	&	4608	&	0.0131	\\
2014.679	&	K2	&	2456906.1439	$\pm$	0.0009	&	4609	&	0.0143	\\
2014.683	&	K2	&	2456907.3743	$\pm$	0.0007	&	4610	&	0.0142	\\
2014.686	&	K2	&	2456908.6017	$\pm$	0.0007	&	4611	&	0.0110	\\
2014.689	&	K2	&	2456909.8351	$\pm$	0.0009	&	4612	&	0.0139	\\
2014.693	&	K2	&	2456911.0670	$\pm$	0.0009	&	4613	&	0.0152	\\
2014.696	&	K2	&	2456912.2954	$\pm$	0.0010	&	4614	&	0.0131	\\
2014.700	&	K2	&	2456913.5182	$\pm$	0.0011	&	4615	&	0.0053	\\
2014.703	&	K2	&	2456914.7501	$\pm$	0.0011	&	4616	&	0.0067	\\
2014.706	&	K2	&	2456915.9801	$\pm$	0.0012	&	4617	&	0.0061	\\
2014.710	&	K2	&	2456917.2124	$\pm$	0.0017	&	4618	&	0.0079	\\
2014.713	&	K2	&	2456918.4442	$\pm$	0.0015	&	4619	&	0.0091	\\
2014.716	&	K2	&	2456919.6708	$\pm$	0.0017	&	4620	&	0.0052	\\
2014.720	&	K2	&	2456920.9031	$\pm$	0.0012	&	4621	&	0.0069	\\
2014.723	&	K2	&	2456922.1363	$\pm$	0.0015	&	4622	&	0.0096	\\
2014.727	&	K2	&	2456923.3678	$\pm$	0.0012	&	4623	&	0.0106	\\
2014.733	&	K2	&	2456925.8276	$\pm$	0.0012	&	4625	&	0.0093	\\
2014.737	&	K2	&	2456927.0565	$\pm$	0.0015	&	4626	&	0.0076	\\
2014.740	&	K2	&	2456928.2893	$\pm$	0.0014	&	4627	&	0.0099	\\
2014.743	&	K2	&	2456929.5245	$\pm$	0.0014	&	4628	&	0.0145	\\
2014.747	&	K2	&	2456930.7498	$\pm$	0.0013	&	4629	&	0.0093	\\
2014.750	&	K2	&	2456931.9823	$\pm$	0.0013	&	4630	&	0.0112	\\
2014.753	&	K2	&	2456933.2111	$\pm$	0.0014	&	4631	&	0.0095	\\
2014.760	&	K2	&	2456935.6742	$\pm$	0.0021	&	4633	&	0.0115	\\
2014.764	&	K2	&	2456936.9083	$\pm$	0.0013	&	4634	&	0.0150	\\
2014.767	&	K2	&	2456938.1387	$\pm$	0.0014	&	4635	&	0.0149	\\
2014.770	&	K2	&	2456939.3644	$\pm$	0.0010	&	4636	&	0.0100	\\
2014.774	&	K2	&	2456940.6075	$\pm$	0.0010	&	4637	&	0.0226	\\
2014.777	&	K2	&	2456941.8305	$\pm$	0.0009	&	4638	&	0.0150	\\
2014.780	&	K2	&	2456943.0654	$\pm$	0.0007	&	4639	&	0.0194	\\
2014.784	&	K2	&	2456944.2902	$\pm$	0.0009	&	4640	&	0.0137	\\
2014.787	&	K2	&	2456945.5247	$\pm$	0.0016	&	4641	&	0.0176	\\
2014.791	&	K2	&	2456946.7497	$\pm$	0.0009	&	4642	&	0.0121	\\
2014.794	&	K2	&	2456947.9825	$\pm$	0.0007	&	4643	&	0.0143	\\
2014.797	&	K2	&	2456949.2108	$\pm$	0.0011	&	4644	&	0.0121	\\
2014.801	&	K2	&	2456950.4447	$\pm$	0.0010	&	4645	&	0.0154	\\
2014.807	&	K2	&	2456952.9058	$\pm$	0.0009	&	4647	&	0.0154	\\
2014.811	&	K2	&	2456954.1270	$\pm$	0.0011	&	4648	&	0.0061	\\
2014.814	&	K2	&	2456955.3605	$\pm$	0.0012	&	4649	&	0.0090	\\
2014.818	&	K2	&	2456956.5928	$\pm$	0.0014	&	4650	&	0.0108	\\
2014.821	&	K2	&	2456957.8256	$\pm$	0.0013	&	4651	&	0.0130	\\
2014.824	&	K2	&	2456959.0584	$\pm$	0.0013	&	4652	&	0.0153	\\
2014.828	&	K2	&	2456960.2862	$\pm$	0.0012	&	4653	&	0.0125	\\
2014.831	&	K2	&	2456961.5165	$\pm$	0.0012	&	4654	&	0.0123	\\
2014.834	&	K2	&	2456962.7431	$\pm$	0.0015	&	4655	&	0.0083	\\
2014.838	&	K2	&	2456963.9825	$\pm$	0.0012	&	4656	&	0.0172	\\
2014.841	&	K2	&	2456965.2100	$\pm$	0.0010	&	4657	&	0.0142	\\
2014.844	&	K2	&	2456966.4425	$\pm$	0.0010	&	4658	&	0.0161	\\
2014.848	&	K2	&	2456967.6703	$\pm$	0.0010	&	4659	&	0.0134	\\
2014.851	&	K2	&	2456968.8950	$\pm$	0.0011	&	4660	&	0.0075	\\
2014.855	&	K2	&	2456970.1334	$\pm$	0.0008	&	4661	&	0.0154	\\
2014.858	&	K2	&	2456971.3649	$\pm$	0.0010	&	4662	&	0.0163	\\
2016.556	&	CT 1.3-m	&	2457591.5515	$\pm$	0.0006	&	5166	&	0.0073	\\
2016.600	&	CT 1.3-m	&	2457607.5433	$\pm$	0.0007	&	5179	&	0.0019	\\
2017.381	&	Pan	&	2457893.0323	$\pm$	0.0011	&	5411	&	0.0041	\\
2018.459	&	ZTF	&	2458286.8041	$\pm$	0.0238	&	5731	&	0.0008	\\
2019.298	&	Myers	&	2458593.2114	$\pm$	0.0031	&	5980	&	0.0019	\\
2019.369	&	Myers	&	2458619.0647	$\pm$	0.0011	&	6001	&	0.0138	\\
2019.386	&	Myers	&	2458625.2146	$\pm$	0.0028	&	6006	&	0.0109	\\
2019.399	&	Myers	&	2458630.1436	$\pm$	0.0007	&	6010	&	0.0177	\\
2019.426	&	Myers	&	2458639.9845	$\pm$	0.0027	&	6018	&	0.0143	\\
2019.470	&	Myers	&	2458655.9785	$\pm$	0.0014	&	6031	&	0.0111	\\
2019.501	&	Myers	&	2458667.0601	$\pm$	0.0009	&	6040	&	0.0178	\\
2019.524	&	Cooney	&	2458675.6711	$\pm$	0.0022	&	6047	&	0.0150	\\
2019.655	&	ZTF	&	2458723.6409	$\pm$	0.0101	&	6086	&	-0.0065	\\
2020.218	&	Myers	&	2458929.1638	$\pm$	0.0016	&	6253	&	0.0150	\\
2020.262	&	Myers	&	2458945.1620	$\pm$	0.0010	&	6266	&	0.0161	\\
2020.292	&	Myers	&	2458956.2361	$\pm$	0.0011	&	6275	&	0.0153	\\
2020.380	&	Myers	&	2458988.2365	$\pm$	0.0014	&	6301	&	0.0215	\\
2020.403	&	ZTF	&	2458996.8593	$\pm$	0.0165	&	6308	&	0.0304	\\
2020.447	&	Cooney	&	2459012.8399	$\pm$	0.0018	&	6321	&	0.0139	\\
2020.494	&	Myers	&	2459030.0747	$\pm$	0.0012	&	6335	&	0.0211	\\
2020.595	&	Myers	&	2459066.9823	$\pm$	0.0026	&	6365	&	0.0123	\\
2020.697	&	Myers	&	2459103.9030	$\pm$	0.0029	&	6395	&	0.0166	\\
2021.108	&	ZTF	&	2459254.0162	$\pm$	0.0074	&	6517	&	0.0030	\\
2021.286	&	Myers	&	2459319.2452	$\pm$	0.0014	&	6570	&	0.0130	\\
2021.300	&	Myers	&	2459324.1680	$\pm$	0.0010	&	6574	&	0.0137	\\
2021.313	&	Myers	&	2459329.0971	$\pm$	0.0017	&	6578	&	0.0206	\\
2021.414	&	Myers	&	2459366.0111	$\pm$	0.0010	&	6608	&	0.0182	\\
2021.498	&	ZTF	&	2459396.7721	$\pm$	0.0095	&	6633	&	0.0155	\\
2021.515	&	Myers	&	2459402.9262	$\pm$	0.0016	&	6638	&	0.0168	\\
2021.576	&	Myers	&	2459425.0707	$\pm$	0.0012	&	6656	&	0.0115	\\
2021.633	&	Myers	&	2459446.0010	$\pm$	0.0023	&	6673	&	0.0225	\\
2022.206	&	Myers	&	2459655.1917	$\pm$	0.0011	&	6843	&	0.0202	\\
2022.304	&	ZTF	&	2459690.8557	$\pm$	0.0105	&	6872	&	-0.0016	\\
2022.462	&	AAVSO	&	2459748.6998	$\pm$	0.0008	&	6919	&	0.0068	\\
2022.472	&	AAVSO	&	2459752.4105	$\pm$	0.0008	&	6922	&	0.0258	\\
2022.475	&	AAVSO	&	2459753.6193	$\pm$	0.0011	&	6923	&	0.0041	\\
2022.479	&	AAVSO	&	2459754.8615	$\pm$	0.0018	&	6924	&	0.0157	\\
2022.486	&	AAVSO	&	2459757.3496	$\pm$	0.0017	&	6926	&	0.0427	\\
2022.499	&	AAVSO	&	2459762.2168	$\pm$	0.0008	&	6930	&	-0.0123	\\
2022.502	&	AAVSO	&	2459763.4568	$\pm$	0.0008	&	6931	&	-0.0028	\\
2022.506	&	AAVSO	&	2459764.6966	$\pm$	0.0008	&	6932	&	0.0064	\\
2022.516	&	AAVSO	&	2459768.3821	$\pm$	0.0012	&	6935	&	0.0003	\\
2022.563	&	AAVSO	&	2459785.6526	$\pm$	0.0032	&	6949	&	0.0431	\\
2022.583	&	Myers	&	2459793.0068	$\pm$	0.0006	&	6955	&	0.0141	\\
2022.593	&	AAVSO	&	2459796.6944	$\pm$	0.0012	&	6958	&	0.0100	\\
2022.627	&	Myers	&	2459808.9980	$\pm$	0.0027	&	6968	&	0.0082	\\
2023.244	&	Myers	&	2460034.1978	$\pm$	0.0006	&	7151	&	0.0179	\\
2023.257	&	Myers	&	2460039.1231	$\pm$	0.0024	&	7155	&	0.0210	\\
2023.388	&	Myers	&	2460087.1137	$\pm$	0.0024	&	7194	&	0.0202	\\
2023.398	&	ZTF	&	2460090.8018	$\pm$	0.0135	&	7197	&	0.0167	\\
2023.402	&	Myers	&	2460092.0371	$\pm$	0.0012	&	7198	&	0.0215	\\
2023.415	&	Myers	&	2460096.9644	$\pm$	0.0029	&	7202	&	0.0266	\\
2023.463	&	Myers	&	2460114.1882	$\pm$	0.0007	&	7216	&	0.0227	\\
2023.476	&	Myers	&	2460119.1039	$\pm$	0.0014	&	7220	&	0.0162	\\
2023.533	&	Myers	&	2460140.0229	$\pm$	0.0009	&	7237	&	0.0159	\\
2023.634	&	Myers	&	2460176.9436	$\pm$	0.0009	&	7267	&	0.0202	\\
2023.678	&	Myers	&	2460192.9427	$\pm$	0.0012	&	7280	&	0.0222	\\
2024.194	&	Myers	&	2460381.2197	$\pm$	0.0009	&	7433	&	0.0255	\\
2024.281	&	Myers	&	2460413.2132	$\pm$	0.0009	&	7459	&	0.0248	\\
2024.369	&	Myers	&	2460445.2055	$\pm$	0.0005	&	7485	&	0.0229	\\
2024.382	&	Myers	&	2460450.1277	$\pm$	0.0012	&	7489	&	0.0229	\\
2024.514	&	Myers	&	2460498.1177	$\pm$	0.0007	&	7528	&	0.0216	\\
2024.584	&	Myers	&	2460523.9547	$\pm$	0.0006	&	7549	&	0.0171	\\
2024.642	&	Myers	&	2460544.8788	$\pm$	0.0005	&	7566	&	0.0219	\\
2024.672	&	Myers	&	2460555.9507	$\pm$	0.0018	&	7575	&	0.0189	\\
2025.262	&	Myers	&	2460771.2892	$\pm$	0.0015	&	7750	&	0.0116	\\
2025.275	&	TESS 	&	2460776.2135	$\pm$	0.0026	&	7754	&	0.0137	\\
2025.275	&	Myers	&	2460776.2210	$\pm$	0.0016	&	7754	&	0.0212	\\
2025.278	&	TESS 	&	2460777.4578	$\pm$	0.0036	&	7755	&	0.0275	\\
2025.282	&	TESS 	&	2460778.6865	$\pm$	0.0028	&	7756	&	0.0257	\\
2025.285	&	TESS 	&	2460779.9243	$\pm$	0.0029	&	7757	&	0.0329	\\
2025.289	&	TESS 	&	2460781.1423	$\pm$	0.0030	&	7758	&	0.0204	\\
2025.312	&	TESS 	&	2460789.7662	$\pm$	0.0028	&	7765	&	0.0304	\\
2025.316	&	TESS 	&	2460790.9871	$\pm$	0.0031	&	7766	&	0.0208	\\
2025.319	&	TESS 	&	2460792.2128	$\pm$	0.0053	&	7767	&	0.0159	\\
2025.322	&	TESS 	&	2460793.4568	$\pm$	0.0034	&	7768	&	0.0294	\\
2025.326	&	TESS 	&	2460794.6873	$\pm$	0.0032	&	7769	&	0.0293	\\
2025.329	&	TESS 	&	2460795.9069	$\pm$	0.0034	&	7770	&	0.0184	\\
2025.332	&	Myers	&	2460797.1435	$\pm$	0.0022	&	7771	&	0.0245	\\
\enddata	
\end{deluxetable}

Figure 2 shows five eclipses with their fitted parabolic minima.  These light curves have large scatter, so that the data are poorly represented by any smooth curve.  Nevertheless, $TESS$ has over 50 flux measures in eclipse, and this number helps substantially for keeping the timing uncertainty within useful values.  The formal measurement error bars range from 0.0026 to 0.0053 days (225--458 seconds).

The total uncertainty in the measured times of minima ($\sigma_{\rm total}$) come from two sources.  The first source of error is the usual result of ordinary noise (including Poisson noise) in the individual measured magnitudes, $\sigma_{\rm meas}$.  Large error bars and sparse data result in relatively large $\sigma_{\rm meas}$, while finely-sampled light curves with high statistical accuracy (like for the the {\it K2} measures) can make for times as good as $\pm$0.0007 days.  The measurement uncertainty in the time of minimum light can be exactly derived by the usual chi-square formalism in a parabola fit.  This measurement error is that which is always quoted for each individual eclipse time.  The second source of uncertainty ($\sigma_{\rm flicker}$) is that the star flickers randomly on all time scales, distorting the light curve and shifting the apparent minimum away from the time of conjunction.  That is, if a random flicker happens to brighten the star during the {\it ingress/egress}, then the observed time of minimum light will be seen substantially {\it after/before} the true time of conjunction.  This source of error is universal, ubiquitous, and it cannot be recovered from in any way.  For U Sco in particular, with the {\it K2} times, Schaefer (2021) derived that the 1-sigma error bars from flickering is $\sigma_{\rm flicker}$=360/$\sqrt{2}$ seconds, for 255 seconds (4.2 minutes or 0.0029 days).  This systematic uncertainty is applicable to each and every eclipse time for U Sco, with large or small or space telescope.  To get the real total 1-sigma error bar ($\sigma_{\rm total}$) for each and every eclipse, the calculated measurement error bar must be added in quadrature to this $\pm$0.0029 days, with $\sigma_{\rm total}^2$=$\sigma_{\rm meas}^2$+$\sigma_{\rm flicker}^2$.  This total uncertainty is the quantity that must be used for $O-C$ curve fitting.

This result has a variety of important implications for observers.  One implication is that the much-vaunted big-telescope and space-telescope measures have no significant advantage over small-telescope ground-based measures.  To be specific, from the eclipse times in Schaefer (2022a), the {\it K2} data have a median $\sigma_{\rm meas}$ of 0.0011 days, while our (Myers) median $\sigma_{\rm meas}$ is 0.0014 days, so the {\it K2} measures are not much better than the small amateur telescope results.  And for what matters, $\sigma_{\rm total}$ is $\pm$0.0031 days for the space-telescope and $\pm$0.0032 days for the small ground-based telescope.  So for $O-C$ concerns, the vaunted satellite photometry missions are equal in eclipse timing accuracy to results from small amateur telescopes.

The eclipses measured while in eruption (starting around day $+$12) are valid measures of the times of the conjunction, yet they will have a substantially larger time jitter due to intrinsic variability than in quiescence.  During the last half of the eruption, the reforming accretion disk has a complex fast-changing structure featuring large gas clouds high above the outer edge of the accretion disk.  These clouds are what make for the eclipses of the central bright spot at all orbital phases.  Around the time of conjunction, the companion star will be eclipsing both the hot inner region of the re-forming disk as well as the bright outer regions of the disk.  We know from the fast changing anomalous eclipses that the outer edges of the disk will have random asymmetries, with these making for asymmetric eclipses.  The variations in eclipse shape will make for an introduced variance in eclipse times.  This variance must be larger than for that introduced by flickering ($\pm$0.0029 days), and we can only know to measure this quantity empirically.  The RMS variation of the $O-C$ is $\pm$0.0101 days for the 3 eclipses during the 1999 eruption, $\pm$0.0084 days for the 12 eclipses during the 2010 eruption, and $\pm$0.0170 days for the 12 eclipses during the 2022 eruption.  This is like the case for $\sigma_{\rm flicker}$ being near 0.0120 days.  The eclipses during eruption provide valid measures of the conjunction times, and they should be used with the appropriate $\sigma_{\rm total}$.  For U Sco in 2022, the 12 eclipse times in eruption provide a measure of the conjunction time with an accuracy near $\pm$0.0120/$\sqrt{12}$, or $\pm$0.0035.  This is useful information that must be included into the $O-C$ fits.

For measuring the period changes, the accuracy depends on the number of measured eclipses and the time distribution of those eclipses.  {\it K2} and {\it TESS} both measure many eclipse times, each with good accuracy, but only over a one-two month interval.  Averaged together for each satellite, we only have one very-well measured point on the $O-C$ curve.  To beat down the flickering and statistical noise in one observing season, we have diminishing returns after a handful of eclipses are measured.  To take a specific example of the year 2025, the 11 {\it TESS} measures have a combined measurement error bar of $\pm$0.0033/$\sqrt{11}$ or $\pm$0.0010 days, while our 3 measures have a combined measurement error bar of $\pm$0.0018/$\sqrt{3}$ or $\pm$0.0010 days.  For measuring $O-C$ curve changes, our 3 short time series from early 2025 are the equal in accuracy and utility as the entire lifetime output from {\it TESS}.

To measure $\dot{P}$ or $\Delta P$, we must have at least three eclipse times widely spaced years apart.  {\it K2} and {\it TESS} can each provide only one seasonal average for the $O-C$ curve.  So {\it K2} and {\it TESS} data are largely useless for the purposes of this paper, where they can only provide a small supplement to the regular ground-based monitoring.  The period-change science comes almost entirely from the small telescope measures at Cerro Tololo (1989--2016) and Siding Spring (2019--2025).

As additions to the 167 eclipse times in Schaefer (2022a), we here report 10 eclipse times during the 2022 eruption that are recorded in the AAVSO database, 23 eclipse times observed by Myers (2022--2025), 7 poor eclipse times from ZTF (2018--2023), one isolated eclipse in 2022 observed by J. Hambsch, and 11 eclipse times from {\it TESS} in 2025.  With these 52 new eclipse measures, we now have 218 eclipse times in total, from 1945 to 2025.  These are presented in Table 1.

The eclipse times are presented in Heliocentric Julian Dates (HJDs) for the ground-based measures and the Barycentric Julian Dates (BJDs) for {\it K2} and {\it TESS}.  The differences between these two reference frames are negligible for the application to U Sco.  The particular eclipses are identified by the number of orbits ($N$, an integer) from a fiducial eclipse.  As in our previous papers, the fiducial ephemeris is $T_{\rm eph}$=2451234.5387+1.23054695$N$.  The observed time for the Nth eclipse is $T$, so the $O-C$ value is $T$-$T_{\rm eph}$.  The $O-C$ curve is a plot of this versus time.  The observed shape of this curve describes the period changes.  If $P$ is constant and the ephemeris is good, then the $O-C$ will be near zero for all times.  If the period is different from the ephemeris period, then the line will have a slope, where the instantaneous slope provides a measure of the period.  If the period changes steadily at a rate of $\dot{P}$, then the $O-C$ curve will be a parabola.  This is the expected general case, where a concave-up parabola indicates a steadily increasing $P$, and a concave-down parabola indicates a steadily decreasing $P$.  The units of $\dot{P}$ are seconds/second, or days/day, or dimensionless.  For a model of the $O-C$ curve, this parabola can be represented as $T_{\rm model}$-$T_{\rm eph}$, where $T_{\rm model}$=$E_0$+$P$$N$+0.5$P$$\dot{P}$$N^2$.  The $T$ data can be fitted to this model equation as a normal chi-square minimization so as to derive the best $E_0$, $P$, and $\dot{P}$.  At the time of each nova eruption, the $P$ will change quickly by a value $\Delta P$, which equals $P_{\rm after}$-$P_{\rm before}$.  In the $O-C$ diagram, a sudden period change appears as a sharp kink, with an upturn indicating a period increase, and with a downturn indicating a period decrease.  With the expected $\dot{P}$ between eruptions and the expected $\Delta P$ for each eruption, the observed $O-C$ curve should appear as a sequence of parabolas connected at the time of each eruption.  A primary goal of this paper is to measure both $\Delta P$ and $\dot{P}$ for and between each eruption.

The $N$ and $O-C$ values for each observed eclipse are presented in Table 1.  For fitting the $O-C$ curve, each individual eclipse is used.  For purposes of display in $O-C$ plots, the timings for each observing season are combined together as a weighted average, where the $\sigma_{\rm tot}$ values are used.  These $O-C$ plots are given in Figures 5 and 6.

\begin{figure}
\epsscale{1.17}
\plotone{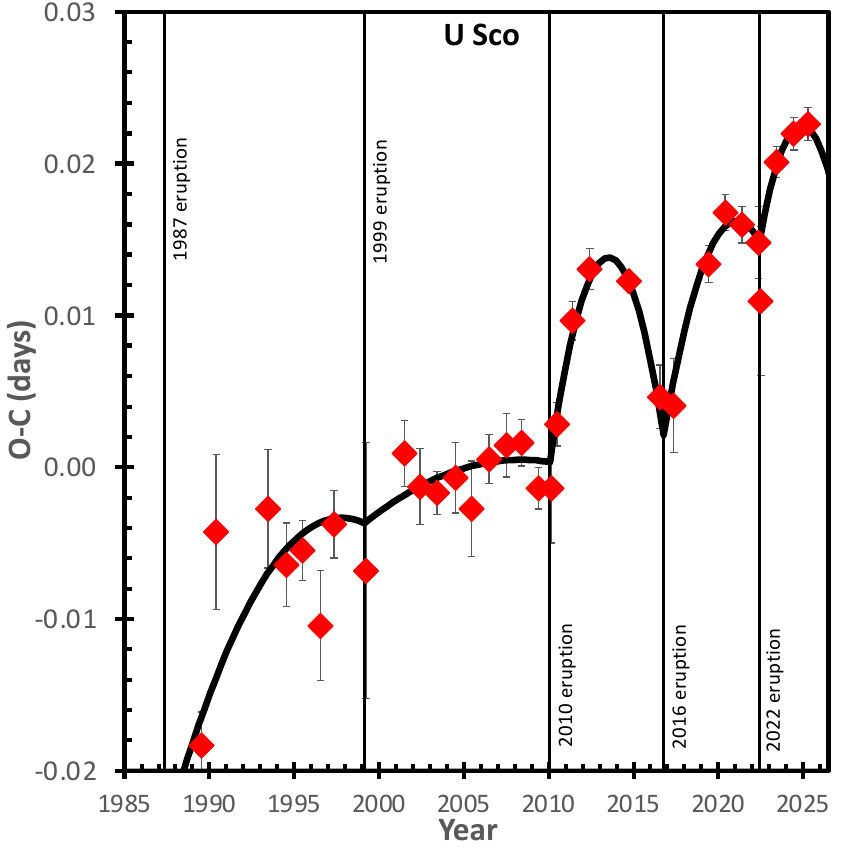}
\caption{U Sco $O-C$ curve for 218 eclipse times.  For display purposes only, the plotted red diamonds are seasonal averages.  The fiducial ephemeris for creating this $O-C$ curve uses a period of 1.23054695 days, and an epoch for $N=0$ of HJD 2451234.5387.  The thin vertical lines indicate the date of an eruption, with the year labeled.  The $O-C$ curve must consist of parabolas of steady period change $\dot{P}$ between eruptions, with kinks at the times of eruption due to the sudden period changes $\Delta P$.  The best fitting model of broken parabolas is represented by the thick black curve.  Startlingly, the $\Delta P$ is seen to change greatly from eruption-to-eruption, and the $\dot{P}$ curvature is also seen to vary greatly from eruption to eruption.  We are not aware of any suggestion or understanding for how some mechanism makes for such changes in the nature of the period-changes.  For the main purpose of this paper, we see and quantify the sharp upward kinks, where the large-positive $\Delta P$ measures are possible only for a huge $M_{\rm ejecta}$.  These ejecta masses are up to 45$\times$ greater than the masses accreted onto the white dwarf in the prior part of the eruption cycle, so the white dwarf must be net losing large masses over each eruption cycle.  With the U Sco white dwarf losing mass over the years, U Sco cannot be a Type Ia supernova progenitor.  }
\end{figure}

\begin{figure}
\epsscale{1.17}
\plotone{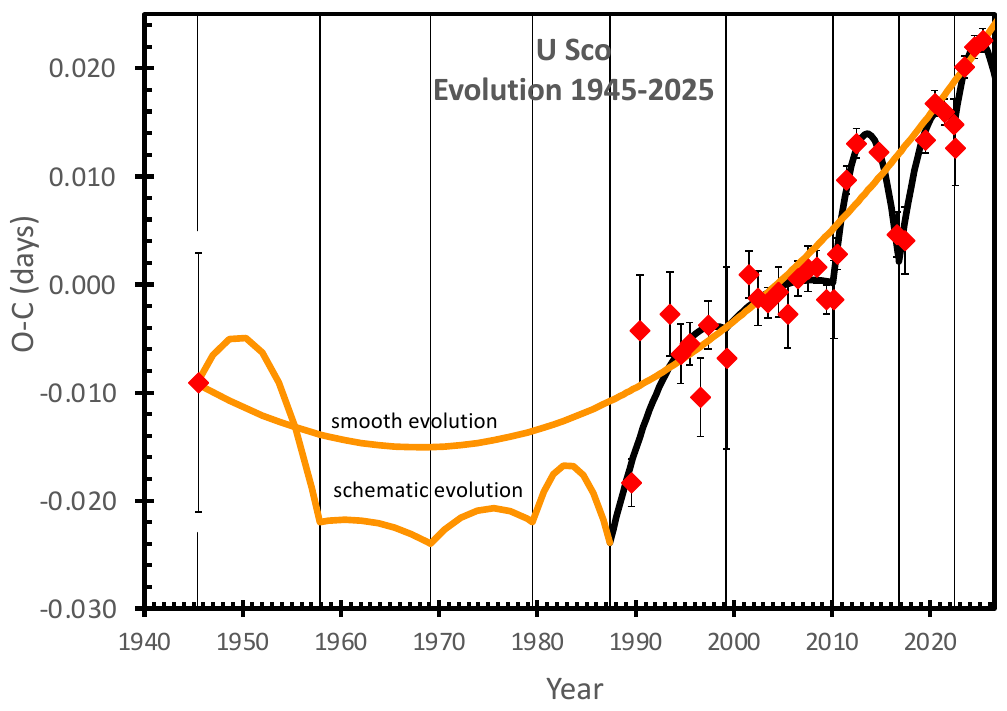}
\caption{U Sco evolution 1945--2025.  The details of this $O-C$ curve are the same as for Figure 5, but here we have added the one observed eclipse time from 1945.  Suddenly, we see that the average evolution from 1945--1989 changes from being flat in this $O-C$ curve to the 1989--2025 case where the curve trends upward (with various jitters superposed).  This trend is represented by the orange unbroken parabola labeled as `smooth evolution', for which the $\dot{P}$ is $+$0.26$\times$10$^{-9}$.  A parabola shape in an $O-C$ curve is for a steady period change, with this $\dot{P}$ both driving and describing the long-term evolution of the system.  That is, the 80 year evolution of U Sco has the orbital period {\it increasing} over time.  U Sco certainly has jitters from 1945 to 1989, with a schematic possibility represented by the connected orange broken-parabolas.  Whether we look at the smooth evolution or the schematic evolution, they all have to come back to the the same 1945 eclipse time, so the overall evolution in 80 years is measured to be significantly increasing in period.  This is in stark contrast to the general expectation that all novae are evolving to shorter-and-shorter $P$.   }
\end{figure}

\section{Period Changes, $\Delta P$ and $\dot{P}$}

\begin{longrotatetable}
\begin{deluxetable*}{lrrrrrrr}
\tablecaption{Period changes for each U Sco eruption}
\tablewidth{600pt}
\tabletypesize{\scriptsize}
\tablehead{
   &   2022.4  &  2016.78$\pm$0.10  &  2010.1  &  1999.2  &  1987.4  & 1979.5	&	1969.1
}
\startdata
N of eruption start	&	6910	&	5233	&	3243	&	0	&	-3499	&	...	&	...	\\
$P_{\rm before}$ just before eruption (days)	&	1.2305401	&	1.2305222	&	1.2305461	&	1.2305456	&	...	&	...	&	...	\\
$E_0$ of eruption start (HJD)	&	2459737.6330	&	2457673.9930	&	2455225.2026	&	2451234.5346	&	2446928.8310	&	...	&	...	\\
	&	$\pm$ 0.0016	&	$\pm$ 0.0021	&	$\pm$ 0.0013	&	$\pm$ 0.0032	&	$\pm$ 0.0022	&		&		\\
$P_{\rm after}$ just after eruption (days)	&	1.2305677	&	1.2305690	&	1.2305737	&	1.2305504	&	1.2305596	&	...	&	...	\\
	&	$\pm$ 0.0000064	&	$\pm$ 0.0000017	&	$\pm$ 0.0000025	&	$\pm$ 0.0000031	&	$\pm$ 0.0000038	&		&		\\
$\Delta P$ of eruption (days)	&	0.0000276	&	0.0000468	&	0.0000276	&	0.0000048	&	...	&	...	&	...	\\
$\Delta P$/P (ppm)	&	$+$22.4 $\pm$ 6.1	&	$+$38.1 $\pm$ 7.6	&	$+$22.4 $\pm$ 1.0	&	$+$3.9 $\pm$ 6.1	&	...	&	...	&	...	\\
Inter-eruption interval	&	2022.4--2025.4	&	2016.78--2022.4	&	2010.1--2016.78	&	1999.2--2010.1	&	1987.4--1999.2	&	1979.5--1987.4	&	1969.1--1979.5	\\
$\tau_{\rm rec}$ (years)	&	$>$3.0	&	5.6	&	6.7	&	10.9	&	11.8	&	7.9	&	10.4	\\
$\dot{P}$ following eruption ($10^{-9}$)	&	 $-$28.1 $\pm$ 10.8	&	 $-$17.2 $\pm$ 2.9	&	 $-$25.9 $\pm$ 3.2	&	 $-$1.3 $\pm$ 1.1	&	 $-$4.0 $\pm$ 1.9	&	...	&	...	\\
Number of magnitudes	&	45	&	329	&	72	&	221	&	26	&	4	&	2	\\
$\langle V \rangle$ (mag)	&	18.01	&	18.11	&	17.78	&	17.97	&	17.98	&	17.82	&	17.90	\\
$M_{\rm trigger}$/$C$ (years)	&	35 so far	&	59 $\pm$ 3	&	122 $\pm$ 8	&	124 $\pm$ 8	&	128 $\pm$ 28	&	124 $\pm$ 8	&	124 $\pm$ 8	\\
$M_{\rm ejecta}$ ($10^{-6}$ $M_{\odot}$)	&	26.5 $\pm$ 7.5	&	45.1 $\pm$ 9.7	&	26.5 $\pm$ 2.4	&	4.6 $\pm$ 7.2	&	...	&	...	&	...	\\
\enddata
	
\end{deluxetable*}
\end{longrotatetable}

The eclipse times in Table 1 can be used in a standard analysis to fit the observed $O-C$ values to a model of the period changes.  For the expected and observed kinks from $\Delta P$ and curves from $\dot{P}$, the model function is a broken parabola, with the breaks at the time of each eruption.  The fit of this model to the observed $O-C$ is made with the usual chi-square minimization for all 218 individual eclipse times.  Well, the 1945 eclipse time is useful only for seeing the general trend of the $O-C$ before 1989, but this is valuable as our best measure of the long-term behavior of the orbit.  The best fitting parameters for each eruption and each inter-eruption interval are tabulated in Table 2.  The five most recent broken parabolas are plotted in Figure 5.  Also placed in Table 2 are the $\langle V \rangle$-band magnitude averaged over the inter-eruption interval, taken from Schaefer (2022a) and updated with the AAVSO light curve.  Also presented are values of $M_{\rm trigger}/C$, calculated as in Schaefer (2022a), which is a cumulative measure of flux scaled to be proportional to the mass accreted by the end of the inter-eruption interval.

The observed $O-C$ curve is stunning for its unexpected complexity and variability.  For the curvature between eruptions, U Sco goes from a near-zero $\dot{P}$ from 1999.2--2010.1 ($+$3.9 $\pm$ 6.1) to an extremely negative $\dot{P}$ from 2010.1--2016.78 ($+$22.4 $\pm$ 1.0).  These measures are from many eclipses and there is no chance of measurement or systematic errors to explain this away.  What happened during the 2010.1 eruption so as to make for huge changes in $\dot{P}$ over the next many years?  The 2010.1 eruption was identical photometrically and spectroscopically to all the other eruptions, so why the big change?  The average quiescent magnitude before-2010.1 and after-2010.1 are similar, so why the big change?

A similar conundrum arises from the kinks in Figure 5.  The $\Delta P$ in 1999 is consistent with zero, but the next three eruptions have huge $\Delta P$ kinks.  What happened in the 1999 eruption that is greatly different from the later eruptions?  Whatever this mechanism forcing large changes in $\Delta P$ from eruption-to-eruption, it has not left a measured difference in the detailed photometry or spectroscopy of the eruptions.  How can the period-change mechanism be hidden?

For information on the mechanisms, we can seek correlations between $\tau_{\rm rec}$, $\dot{P}$, and $\langle V \rangle$.  We expect that $\dot{P}$ and $\langle V \rangle$ might be correlated (just as for T Pyx, see Schaefer 2023b) with a mechanism that has a high accretion rate (as evidenced by a bright $\langle V \rangle$) causing a high $\dot{P}$.  But in Table 2, we see no correlation at all.  Further, we might expect a correlation between $\langle V \rangle$ and the time between eruptions, with the basis that a high accretion rate (with a bright quiescence) should mean a short time interval between nova events.  But again, Table 2 shows no such  correlation.  With the brightness in quiescence not being correlated with $\dot{P}$ or $\tau_{\rm rec}$, we can then have no confidence in the method of {\it predicting} upcoming eruptions as based on the average brightness in quiescence.

Another attempt to find a new and useful correlation in Table 2 is to see if $\dot{P}$ varies with $\Delta P/P$.  We have four instances of $\Delta P/P$ and the $\dot{P}$ {\it following} the eruption, and these appear correlated.  That is, the three eruptions with large $\Delta P/P$ are followed by intervals of large-negative $\dot{P}$, and the one eruption with a near-zero $\Delta P/P$ is followed by a near-zero $\dot{P}$.  We do not know what mechanism during an eruption would set $\Delta P/P$ as well as set $\dot{P}$ for the next decade.  The existence of this correlation is not convincing, mainly because it is only based on the values from just one eruption being different from the three others.  So we can only point to this possible correlation as being `suggestive'.  We must await perhaps two decades of eclipse timings to measure another pair of points to test this possible correlation.

Figure 5 shows a complex dance of period changes over the last 38 years.  We see that the {\it negative} $\dot{P}$ effects are largely offset by the {\it positive} $\Delta P$ effects.  That is, from 1987--2025, the $O-C$ curve appears to be jittering around a straight line (running from the lower-left to the upper-right in Fig. 5), which makes for a largely-unchanging smoothed period.  We know of no mechanism that can regulate the two greatly different effects to keep them comparable in magnitude yet with opposite signs.  This warns us that quantifying the U Sco evolution with only $\dot{P}$ or $\Delta P$ alone will give the wrong answer.  What we really should be looking at for the evolutionary period changes is the sum over many eruption cycles of $\dot{P}$+$\Delta P$/$\tau_{\rm rec}$.  All theoretical models for the general evolution of cataclysmic variables overlook this $\Delta P$/$\tau_{\rm rec}$ term.

With the 1945 eclipse observed with the Harvard plates, we can extend Figure 5 from 38 years to 80 years, as shown in Figure 6.  The extension from 1987 back to 1945 is presumably some set of broken parabolas, for which a schematic possible case is shown by the orange curve.  (This schematic evolution includes an undiscovered eruption around 1957, an interval with essential no archival data to allow later discovery.  Even with plentiful archival data, the 60-day U Sco eruption could easily be hidden in the months October to February over which the RN is too close to the Sun for anyone to discover the event.  It seems more likely for U Sco to have an undiscovered eruption $\sim$1957 than for it to have a 23.7 year interval between eruptions.)  The importance of the 1945 eclipse is that its time is far from that expected for any extrapolation of the crude 1989--2025 line.  That is, if U Sco is really following the smoothed linear trend from 1989--2025, then the extrapolation back to 1945 would predict an $O-C$ of something like $-$0.060 days ($-$86 minutes), and this is impossible given the 1945 light curve.  So the 1945 eclipse forces the overall shape of the $O-C$ curve to have a significant upturn.  That is, on the timescale of 80 years, the U Sco period is getting larger and larger.  This trend can be represented by a single parabola, drawn as an orange parabola in Fig. 6.  Superposed on this evolutionary change are the fine year-by-year effects of $\Delta P$ and $\dot{P}$.  So our best measure of the long-term evolution of U Sco is that $P$ is on average {\it increasing} with a $\dot{P}$ of $+$0.26$\times$10$^{-9}$.  This is sharply at odds with the general expectation that CVs and nova must be evolving steadily to shorter-and-shorter periods.

\section{Mass of the Nova Ejecta}

The $M_{\rm ejecta}$ for novae is critical for a variety of important calculations, including the evolution of the binaries, the enrichment of heavy elements in the galactic interstellar medium, and the SNIa progenitor problem.  The long lasting trouble is that it is hard to measure $M_{\rm ejecta}$ with useable accuracy.  Indeed, all the old traditional observational measures have real error bars of 2--3 orders of magnitude.  Further, the current state of theoretical predictions is also where the real uncertainties are 4 orders of magnitude.  This poor state was already widely recognized back even in 1983.  This was a large part of the realization that the decades-long intensive-observing program to measure $\Delta P$ was required as the only method to measure a reliable $M_{\rm ejecta}$.

\subsection{Old Observational Measures}

To estimate the mass of the ejecta ($M_{\rm ejecta}$) in a nova eruption, the old methods have all been based on measuring the flux in one of the hydrogen emission lines.  Unfortunately, this hoary old method has five problems, each with independent orders-of-magnitude uncertainties (Schaefer 2011, Appendix A):  

{\bf A.} The recombination coefficient depends critically on the gas temperature, which is not known.  ``Even for the detailed physical analysis of the 1979 eruption of U Sco, Williams et al. (1981) could only consider temperatures over the range of 10,000--25,000, and the resulting shell masses varied by a factor of 1500.''  

{\bf B.} ``The filling factor,  $\epsilon$, is based entirely on guesses, with no factual basis. No calculations are made based on the theory of turbulence in the shell, and there is no way to measure the filling factor based on line ratios. No account is ever taken of the hollow inside the shell caused by the turn-off of ejection, the inevitable bipolar outflows caused by the accretion disk, or the equatorial plane of enhanced emission (as causes the triple peaked lines observed for U Sco). While the value of $\epsilon$ cannot be gotten from theory or observation, workers can only resort to guesses. For U Sco, published values for $\epsilon$ range from $<$0.001 to 0.1, but I see no reason to think that larger or smaller values are not unreasonable. Even with this, we have an uncertainty of two orders of magnitude in $\epsilon$ and hence also in $M_{\rm ejecta}$.''  

{\bf C.} The old methods have $M_{\rm ejecta}$ depending on the square of the distance $D$.  For prior work, this uncertainty in $D$ made for an order-of-magnitude uncertainty in the ejecta mass.  Fortunately, with the recent analysis of {\it Gaia} parallax plus the full Bayesian priors based on other information, the distance to U Sco has been greatly improved, now with $D$=6260$^{+1170}_{-840}$ parsecs (Schaefer 2022c).  Now, the 1-sigma range only suffers a factor of 1.9$\times$ uncertainty due to the distance.  

{\bf D.} ``The traditional method with line fluxes implicitly assumes that the shell is optically thin, so that we can see all the line photons being emitted. Early during the eruption, the shell is certainly optically thick, and it becomes thin only after substantial expansion. We know in the case of U Sco that the inner binary is hidden behind an optically thick photosphere until about 13 days after the peak, when the eclipses suddenly start and when the supersoft source becomes first visible (Schlegel et al. 2010; Schaefer et al. 2010a). Presumably, the outer volume of the shell can be optically thin, while the inner volume (and the volume it shadows) is not included in the mass estimate. All U Sco estimates are made based on spectra from dates when the shell is still optically thick. A correction factor is needed, and this will be large when the nova is near peak and will then decrease to near unity as the eruption ends. I have seen no consideration or calculation of this correction factor in the literature.''  

{\bf E.} ``The adopted volume of the shell scales as the cube of the expansion velocity, with $v$ taken from some emission- line profile. However, there is no understanding of which velocity to take from the line profile so as to correlate with the outer edge of the shell. Should we take the HWHM or the HWZI of the line profile, or some other velocity? The line widths change substantially with time as the photosphere recedes, so should we take the line width at the nova peak, at the time of the observation, or at some other time? For U Sco, Zwitter \& Munari (2000) give the HWZI to vary as 5015--130$\Delta t$ km s$^{-1}$, while Anupama \& Dewangan (2000) report HWZI values of 5065 and 3262 km s$^{-1}$ for $\Delta t$ values of 0.5 and 12 days. We also expect differences in velocity by perhaps a factor of 2.5 with the angle from the orbital pole due to bipolar ejection (e.g., Walder et al. 2008). All these problems certainly lead to errors in $v$ of a factor of $\gtrsim$2. With this, the resulting error in $M_{\rm ejecta}$ will be a factor of $\gtrsim$8.''

With these five problems, the real uncertainty in all prior published measures of $M_{\rm ejecta}$ is 2--3 {\it orders-of-magnitude}.  Such is useless to answer any of the astrophysics questions.  More generally, the same applies to all novae, where the mass of the ejecta is always largely unknown.  This was well-known even back in 1983 (c.f., Schaefer \& Patterson 1983), and this served as a key reason for starting the entire program of measuring $\Delta P$ for U Sco, RNe, and classical novae in general.

\subsection{Theoretical Models}

Theoretical models of nova explosions should, in principle, be able to predict $M_{\rm ejecta}$.  To this end, theorists have been constructing models for at least four decades.  Recent technical discussions show a pride at which modelers are adding yet more and better nuclear reaction rates and adding the effects of General Relativity.  Effectively, all of these have been 1-dimensional calculations, with such being the best that can be done even to recent years.  The 1-D models might cover much of the basic physics, but the long experience in astrophysics with explosions is that the solutions change substantially when going to 2-D and 3-D.  For nova explosions in particular, critical physics requiring 3-D calculations includes the convection that mixes the underlying WD material with the accretion layer, and the stirring of the nova envelope by the companion star that ejects the majority of the gas.  So we already know that these old-style model predictions of $M_{\rm ejecta}$ are 'incomplete' and `unreliable'.

The large uncertainties in model predictions were recognized early on by simply comparing predicted $M_{\rm ejecta}$ values for specific novae from various groups worldwide.  Schaefer \& Patterson (1983) collected three published measures for HR Del to be 90, 250, and 1500 M$_{\odot}$, with deviations of a factor of 17$\times$.  For U Sco, Schaefer (2011) collected published theory predictions to be 0.21, 1.8, 0.43, and 4.4 in units of $10^{-6}$ M$_{\odot}$, with deviations of a factor of 22$\times$.  For the U Sco case of a 1.35 M$_{\odot}$ ONe WD, Starrfield et al. (2025) predicts ejecta masses from {\it zero} to 0.012 to 0.42 in units of $10^{-6}$ M$_{\odot}$.  That they have published models predicting $M_{\rm ejecta}$=0 shows that their models have no reliability or utility.  Other models for the U Sco case predict 3.4--4.4 (Jos\'{e} \& Hernanz 1998) and 1.9 (Rukeya et al. 2017), in units of 10$^{-6}$ M$_{\odot}$.  The point is that model predictions show stark disagreement between all the practitioners, so we know that we cannot reliably model $M_{\rm ejecta}$ to any accuracy better than 1--2 orders of magnitude.

The problem is actually substantially worse, simply because ordinary choices by modelers make for huge changes.  For models of 1.35 M$_{\odot}$ ONe WDs, just by making a free choice in the manner in which the underlying WD material gets mixed in, Starrfield et al. (2024) reported predicted $M_{\rm ejecta}$ that varied over a range by a factor of 3300$\times$.  This is horrifying because we have no way of knowing this input, the range of physical possibilities is substantially larger than assumed in the models, so the real uncertainty in the predicted $M_{\rm ejecta}$ is at least 3300$\times$ from this one issue alone.  Even worse, for the same model as applicable to U Sco, the simple change in the resolution used by the code (from 95 to 300 mass zones) changes the predicted $M_{\rm ejecta}$ by a factor of 24$\times$.  And the old style models suffer from hysteresis, where the running of models through multiple eruptions has the $M_{\rm ejecta}$/$M_{\rm accreted}$ ratio flopping chaotically between $<$1 and $>$1 (Hillman \& Kashi 2021).  So the real range of predictions has an uncertainty of roughly 4 orders of magnitude, all for unknowable free changes in the model input.  This is to say that the old-style 1-D models have real error bars that extend around 4 orders of magnitude in size.

And it gets worse for evaluating the old-style models.  The trouble is the recent realization that these old theory models have not included the dominant mass ejection mechanism.  So the old-style calculations are largely irrelevant because they are not considering the primary ejection mechanism.

The new dominant mechanism is that the companion star circles within the outer regions of the hot quasi-stationary envelope surrounding the WD (Sparks \& Sion 2021, Shen \& Quataert 2022, Chomiuk, Metzger, \& Shen 2020).  This envelope is what becomes visible as the usual super-soft source.  The motion of the companion stirs up the gas in the envelope, ejecting large quantities of gas.  This motion has the companion star suffering drag forces (called FAML) that lowers the orbital period and sends angular momentum out of the binary with the ejected gas.  Unfortunately, at least for now, the situation of the binary interaction with the hot envelope is beyond current modeling capability with any useful accuracy.  This means that the new-style models have not been made yet, so there is no useful or reliable prediction of $M_{\rm ejecta}$.

In all, neither the old-style 1-D models, nor the new-style binary-interaction models, can return any reliable prediction with an accuracy of better than 4 orders of magnitude.  And from the previous Section, the old traditional observational methods all have real measurement uncertainties of 2--3 orders of magnitude.  This is a bad situation, recognized well before 1983, that has persisted into recent years.  So, from neither theory nor observation do modern astronomers have any reliable or useful idea as to $M_{\rm ejecta}$.  But getting reliable and useful measures of $M_{\rm ejecta}$ retains its importance for many front-line questions.  So some good solution is desperately needed.

\subsection{$\Delta P$ Method}

The confident solution to this problem was realized back in 1983 (Schaefer \& Patterson 1983).  A measure of the $\Delta P$ can be used to derive $M_{\rm ejecta}$.  The physics is from Kepler's Law and the conservation of angular momentum, where the loss of mass from the white dwarf will exactly change the orbital period.  This then has the ejecta mass being measured with a simple timing experiment of high accuracy.  This method does not use any input from the highly-uncertain gas temperature, filling factor, distance, internal shadowing, or shell volume.  This $\Delta P$ method is reliable and of good accuracy.

Unfortunately, this $\Delta P$ method requires decades of pain-staking observations and is applicable to only a small number of novae.  For this method to work for any particular nova, its quiescent counterpart must display some sort of phase marker (like an eclipse or a large amplitude ellipsoidal modulation), and be bright enough to allow frequent time series with modest-sized telescopes.  (For non-recurrent novae, the brightness and phase markers had to be particularly prominent so as to allow their visibility on many archival plates, as the only way to get {\it pre}-eruption times.)  For workable novae, the program is then to take time series on many eclipses per observing season (to beat down the timing uncertainties), season after season, decade after decade.  For classical novae with adequate pre-eruption data extending for decades before the nova event, the observers then must patiently accumulate decades of steady eclipse timings to measure kinks and curvatures in the $O-C$ curve to sufficient accuracy.  This program started with RNe, because they were realized to be the case where we can realize in advance that some particular star will erupt as a nova so as to allow the regular monitoring year-after-year while awaiting the upcoming nova.

This work to get the reliable $M_{\rm ejecta}$ measure can only be run with a decades-long intensive program, such as is unavailable for most workers or for most novae.  So for many novae, workers will publish the only means available to estimate the important ejecta-mass, resulting in hundreds of papers quoting uselessly-bad values derived from Balmer line fluxes.  Unfortunately, many readers do not realize that the real error bars are 2--3 orders-of-magnitude.  Theory calculations for $M_{\rm ejecta}$ are inconsistent with each other by 1--2 orders-of-magnitude for the same nova (Schaefer 2011), so these are useless to address other questions (like for evolution and demographics).  Even constraints on $M_{\rm ejecta}$, for example from theory estimates of the trigger mass, have huge uncertainties of their own, especially from knowing how much ejecta is dredged up from the surface of the exploding white dwarf.  The end result is that the community of nova scientists has been awash with bad measures of $M_{\rm ejecta}$, deceiving us into thinking that we have some idea as to what the mass really is.

The only solution is for a few hard-fought novae with measured $\Delta P$.  The first case was for the bright DQ Her with adequate archival material (Ahnert 1960), but this measure was later proven to be erroneous (Schaefer \& Patterson 1983, Schaefer 2020a), while the final exhaustive measure gave a near-zero $\Delta P$ (Schaefer 2020b).  The first real measure of $\Delta P$ came for the nova BT Mon, which increased its orbital period by 45.2$\pm$0.3 parts-per-million (ppm) after its 1939 eruption (Schaefer \& Patterson 1983).  The fact that BT Mon {\it increased} its $P$ across the nova eruption, making for a slight increase in the binary separation, provided a substantial motivation and justification for the alluring and insightful `Hibernation Model' of the evolution of cataclysmic variables (Shara et al. 1986).  The long-lasting and tedious program of measuring eclipse times for RNe started in 1989, intentionally to address the question of whether $M_{\rm ejecta}$$<$$M_{\rm accreted}$ and whether RNe are the progenitors of Type Ia supernovae.  Results were slow at first, as the long series of eclipse times had to be accumulated.  Around twenty years ago, the RNe program was expanded to start including non-recurrent novae for which adequate archival data were available.  Currently (Schaefer 2020b, 2025a), good $\Delta P$ measures are known for 11 systems (4 RNe and 7 non-recurrent novae), with one system, U Sco, having measures for 3 eruptions.  This paper reports on $\Delta P$ for U Sco for the 2022 eruption.

The basic equation relating $\Delta P$ and $M_{\rm ejecta}$ comes from Kepler's Law (and the conservation of angular momentum).  
\begin{equation}
\Delta P_{\rm ml} = 2  \frac{M_{\rm ejecta}}{M_{\rm comp}+M_{\rm WD}} P.
\end{equation}
The subscript `ml' indicates the effects on $\Delta P$ from the mass loss of the white dwarf resulting from the nova ejecta.  This basic equation has been derived by many workers\footnote{ The factor 2 in Equation 1 has corrections for small effects that can change the factor by up to a few per cent.  The main factor is that the companion will intercept a small fraction of the ejecta, reducing the factor of 2 by 6$\pm$2\% for U Sco (see equations 4--6 of Schaefer 2020a).  Another effect is that the mass ejected from the white dwarf might carry a different angular momentum than the orbital value, but this effect is minuscule because any asymmetry in the ejecta site is averaged out by the rotation and revolution, while the moment arm for any ejecta is small compared to the orbital size. A further effect is that the ejecta hitting the companion has a momentum kick, but this effect is negligibly small even for the case of supernovae.}, and it was old even for Schaefer \& Patterson (1983).  This assumes that the angular momentum carried away by the ejecta is the same as the material originally had on the white dwarf surface.  The values for $P$, $M_{\rm WD}$, and the comparison star mass $M_{\rm comp}$ can be independently know to reasonably good accuracy.  Thus, a simple timing experiment can return the period change and then allow a reliable and accurate calculation of $M_{\rm ejecta}$.

\subsection{New Constraints}

There is only one systematic problem that needs to be addressed, and that is that the binary star can suffer angular momentum loss during the nova event that will also change the $P$.  The overall effect of angular momentum losses can be labeled as $\Delta P_{\rm aml}$.  This loss is entirely from the companion star plowing through gas put out by the white dwarf as part of the eruption.  As the companion moves through the gas, its forward motion is subject to a drag force that slows its velocity, hence decreasing the orbital radius and increasing the orbital period.  The angular momentum robbed from the binary goes into gas ejected from the system.  Critically, angular momentum can only be lost from the binary (after all, there is no outside source to {\it give} angular momentum to the binary), so $\Delta P_{\rm aml}$ must necessarily be $\leq$0.

The $\Delta P_{\rm aml}$ can be broken up into two components.  In olden times, the only recognized source, called `fractional angular momentum loss' or FAML, was for the companion star being dragged through the fast expanding shell of high velocity gas as part of the ejecta.  Livio, Govarie, \& Ritter (1991) gives this effect as 
\begin{equation}
\Delta P_{\rm FAML} = -\frac{3}{4} P \frac{M_{\rm ejecta}}{M_{\rm comp}} \frac{V_{\rm orb}}{V_{\rm shell}} (\frac{R_{\rm comp}}{a})^2.
\end{equation}
Here, the $V$ values are for the velocity of the companion star in its orbit and for the expansion velocity of the shell.  $R_{\rm comp}$ is the radius of the companion star and $a$ is the usual semimajor axis of its orbit.  This term is always negative and is always smaller in magnitude than $\Delta P_{\rm ml}$.  This means that $\Delta P_{\rm ml}$+$\Delta P_{\rm FAML}$ must always be positive.  If these are the only two nova effects, then the observed $\Delta P$ must always be positive.  But this is contradicted by measurement for all non-recurrent novae other than BT Mon (Schaefer 2020a).  This requires that there be some additional angular momentum loss source, as only that can account for the many cases with $\Delta P$$<$0.

In recent years, multiple workers have realized that nova eruptions not only eject fast moving gas, but also puff up a hot envelope (e.g., Shen \& Quataert 2022).  This envelope is what creates the super-soft X-ray source that shines through the gas (after the transition to an optically thin shell) for all novae, and the fast expansion of the shell is launched from its outer layers.  This envelope is quasi-stationary, and comparable in size to the binary orbit.  As the companion star plows through the outer edges of this envelope, the drag will rob angular momentum from the orbit, making for a $\Delta P_{\rm env}$.  The details of the envelope density are not yet well-modeled, and they are sufficiently complex so that any one equation will not be adequate in accuracy.  Nevertheless, we do know some simple parts of the effect.  The $\Delta P_{\rm env}$ will be proportional to the ejected mass.  And the envelope density at the distance of the companion's orbit will vary exponentially as the scale height above the white dwarf.  The lifetime of the envelope (and its drag effect) will be proportional to the nova decline time.
\begin{equation}
\Delta P_{\rm env} \propto t_3 M_{\rm ejecta} e^{-M_{\rm WD}}.
\end{equation}
Critically, $\Delta P_{\rm env}$$\leq$0.  While we cannot now know $\Delta P_{\rm env}$ with any accuracy, the limits and proportionality will allow us to recognize extreme cases where the period change is either near-zero or very-large.

The total effect observed for a nova event is 
\begin{equation}
\Delta P = \Delta P_{\rm ml} +\Delta P_{\rm aml}  =  \Delta P_{\rm ml} +\Delta P_{\rm FAML}  +\Delta P_{\rm env}.
\end{equation}
For ordinary classical novae (e.g., QZ Aur, HR Del, DQ Her, RR Pic, V1017 Sgr), the total effect is negative, so the angular momentum loss dominates over the ejection of mass.  We cannot work out the $M_{\rm ejecta}$ in these cases because we do not know the constant of proportionality in Equation 3.  For the ordinary classical nova with $\Delta P$$>$0 (BT Mon and V445 Pup), we know that the mass loss must be fairly large, so as to overcome the effects of the angular momentum loss.

For RNe, we have a case where the angular momentum loss must be negligibly small.  For $\Delta P_{\rm FAML}$, RNe must always have small $M_{\rm ejecta}$ and are always observed to have large $V_{\rm shell}$.  So, compared to other novae, RNe must have $\Delta P_{\rm FAML}$ small by something like 2 orders-of-magnitude.  For $\Delta P_{\rm env}$, RNe must have near maximal $M_{\rm WD}$, small $M_{\rm ejecta}$, and are observed to have small $t_3$.  So compared to other novae, the overall effect is that RNe have $\Delta P_{\rm env}$ small by something like 2 orders-of-magnitude.  The combined effect is that RNe must have a small $\Delta P_{\rm aml}$.  This effect is so small as to be negligible (in comparison to $\Delta P$), so we are safe to say $\Delta P_{\rm aml}$$\approx$0, or that $\Delta P_{\rm aml}$=0 to all needed accuracy.  

Let us give a numerical calculation to illustrate the case for U Sco.  The observed $\Delta P$/$P$ is 22 ppm, as averaged over our four measured eruptions.  For the average ejecta as calculated below, the effect of mass loss is $\Delta P_{\rm ml}$/$P$ is 22 ppm.  For this average ejecta and the shell velocity given by the FWHM, the FAML effect is $\Delta P_{\rm FAML}$/$P$ equal to $-$0.074 ppm.  From Equation 4, the value for $\Delta P_{\rm env}$/$P$ is near to zero, but this is by construction as the ejecta masses are calculated with a zero value.  There is no ready way to estimate $\Delta P_{\rm env}$, because we have no useable model.  Nevertheless, we can make a crude estimate of the relative size of $\Delta P_{\rm env}$/$P$ for U Sco by using Equation 5 to scale from other classical nova.  Only four normal classical novae have observed $\Delta P$ and observed $t_3$; DQ Her, HR Del, BT Mon, and RR Pic.  Using the binary properties as given in Schaefer (2023b), the U Sco $\Delta P_{\rm env}$/$P$ is factors of 0.005, 0.002, 0.009, and 0.015 times, respectively, smaller than for the four ordinary novae.  With this, $\Delta P_{\rm aml}$=$\Delta P_{\rm FAML}$+$\Delta P_{\rm env}$ is negligibly small for U Sco, and also for all RNe.  Any deviation from `negligibly small' will only make the derived $M_{\rm ejecta}$ larger, so any invocation of any significant AML will only make the ejected mass even larger. 

So, for RNe, $\Delta P$=$\Delta P_{\rm ml}$ to good accuracy.  Then with Equation 1, we have
\begin{equation}
M_{\rm ejecta} = 0.5 (M_{\rm comp}+M_{\rm WD})  \frac{\Delta P}{P},
\end{equation}
for U Sco and RNe.  So there we have it, a reliable way to derive the ejecta mass from the $\Delta P$.  This equation works for RNe.  This physics has no pernicious systematic problems, and no hidden source of large uncertainty.  This is the result that allows for the consummation of our four-decade-long program to measure $M_{\rm ejecta}$.

\subsection{$M_{\rm ejecta}$ For U Sco}

To calculate $M_{\rm ejecta}$, we need the stellar masses in the binary.  The WD mass is accurately known from the Nomoto plot (Shen \& Bildsten 2009, Figure 7), where the ten-year recurrence time scale forces the white dwarf to be close to 1.35$\pm$0.03 $M_{\odot}$.  Further, detailed modeling of U Sco gives 1.36 $M_{\odot}$ (Shara et al. 2018) and 1.37$\pm$0.01 $M_{\odot}$ (Hachisu \& Kato 2018), while radial velocity curves give 1.55$\pm$0.24 $M_{\odot}$ (Thoroughgood et al. 2001).  So we take the mass of the U Sco WD to be 1.36$\pm$0.03 $M_{\odot}$.  The companion mass has not been usefully measured, but the known radius, surface temperature, and subgiant status points to a mass of 1.0$\pm$0.2 $M_{\odot}$.  The value of $M_{\rm comp}+M_{\rm WD}$ is then 2.36$\pm$0.2 $M_{\odot}$, introducing an 8\% error for the $M_{\rm ejecta}$ calculation.

The four derived $M_{\rm ejecta}$ values are presented as the last line of Table 2.  For the last three eruptions, the ejecta weighs in at $>$2.6$\times$10$^{-5}$ $M_{\odot}$.  The 2022.4, 2016.78, 2010.1, and 1999.2 eruptions have ejecta masses of 26.5$\pm$7.5,	45.1$\pm$9.7,	26.5$\pm$2.4, and	4.6$\pm$7.2 in units of $10^{-6}$ $M_{\odot}$.  With our very-long $O-C$ curve, we have measures of $M_{\rm ejecta}$ over four eruptions.  The sum of $M_{\rm ejecta}$ for the four eruptions is (103$\pm$14)$\times$$10^{-6}$ $M_{\odot}$.

Finally, after four decades of intensive work, we can test whether U Sco can be a supernova progenitor.

\section{U Sco is {\it Not} a Supernova Progenitor}

We have two tests for whether U Sco is a Type Ia supernova progenitor.  The first test is whether the ejecta mass is greater than the accreted mass.  The second test is whether U Sco is a neon nova, and hence whether it has a CO white dwarf.

\subsection{Is $M_{\rm ejecta}$ $>$ $M_{\rm accreted}$?}

For U Sco to be a supernova progenitor, its white dwarf mass must be increasing over time.  Schematically, RNe are great candidates as progenitors, because they must contain a near-Chandrasekhar-mass WD that has mass being piled on at near maximal accretion rate.  But this simple case does not account for the mass lost during each nova event.  So the question comes down to a balance between the mass accreted between eruptions ($M_{\rm accreted}$) great-then or less-than the mass ejected by each eruption ($M_{\rm ejecta}$).  If $M_{\rm ejecta}$$>$$M_{\rm accreted}$, then the WD is losing mass over each eruption cycle, and it cannot be a supernova progenitor.  To become a SNIa, $M_{\it WD}$ must be increasing over each eruption cycle.

We now have measures of $M_{\rm ejecta}$, so the task becomes getting measures of $M_{\rm accreted}$.  For this, we have three estimates.  The first estimate is that the accreted gas is coming onto the WD at nearly the maximal rate (i.e., close to $10^{-7}$ $M_{\odot}$ yr$^{-1}$) for an average time of 10 years between eruptions, so $\log[M_{\rm accreted}]$ must be close to -6.0 for units of $M_{\odot}$.  The second estimate is by interpolating in the table of Yaron et al. (2005) for a 1.35 $M_{\odot}$ white dwarf at the same accretion rate, for a needed log of the accretion mass of $-$6.5.  From the Nomoto plot in Shen \& Bildsten (2009), the trigger requirement is $-$5.8 in log units, for a ten-year recurrence time on a 1.35 $M_{\odot}$ white dwarf.  Shen \& Bildsten (2009) compare in detail predicted trigger levels from many workers, finding that the theory results vary with the assumed details by a factor of around 2$\times$ between theorists.  So we take the median $\log[M_{\rm accreted}]$ of $-$6.0 with an uncertainty of something like $\pm$0.3.  Thus, we take $M_{\rm accreted}$ as being within a factor of 2$\times$ of 1$\times$$10^{-6}$ $M_{\odot}$.

Finally, after 4 decades of work on U Sco, we can compare $M_{\rm ejecta}$ and $M_{\rm accreted}$.  For the last three eruption, $M_{\rm ejecta}$$>$$M_{\rm accreted}$ by factors of 26.5$\pm$7.5,	45.1$\pm$9.7,	26.5$\pm$2.4.  The 1999.2 eruption has a sufficiently large error bar so as to not be able to decide the question.  If we sum up over all four measured eruptions, $M_{\rm ejecta}$ is (103$\pm$14)$\times$$10^{-6}$ $M_{\odot}$, while $M_{\rm accreted}$ over four eruption cycles is 4$\times$$10^{-6}$ $M_{\odot}$.  The U Sco white dwarf is ejecting 26$\times$ as much gas as it is accreting over each eruption cycle.  So the WD is being fast whittled down over every eruption cycle.  So $M_{\it WD}$ is certainly decreasing over evolutionary time.  So U Sco cannot be a Type Ia supernova progenitor.

This is the primary result from the program started in 1989.  U Sco is not a supernova progenitor.  There is no opening for any significant uncertainty, and we know of no way to impeach this critical measurement.  (For example, any attempt to add any angular momentum loss mechanism is only {\it increase} $M_{\rm ejecta}$, because there is no outside source that can add angular momentum to the system.)  With the confident rejection of U Sco as a progenitor, the broad class of single-degenerate models takes a big hit, because one of its primary exemplar and prototype is proven wrong.

\subsection{Is U Sco a Neon Nova?}

For a single-degenerate progenitor to become a normal SN{\rm I}a, the WD must have a carbon/oxygen (CO) composition, and must not have an oxygen/neon (ONe) composition.  The reason is that the thermonuclear burning of the oxygen and neon provide relatively small amounts of energy (compared to carbon burning) and cannot provide the energy of a SN{\rm I}a.  So any system with an ONe WD cannot be a supernova progenitor.  This can be tested for CVs that undergo classical nova eruptions, where dredged-up WD core material (rich in neon) is ejected, so as to be visible in the expanding ejecta shell (Starrfield, Sparks, \& Truran 1986).  A nova that displays a high abundance of neon in its ejecta is called a `neon nova'.  The only source for a high abundance of neon in the ejecta is for the eruption to dredge-up material from the core of an ONe WD.  So a CV that has a neon nova eruption must have an ONe WD and hence cannot possibly become a normal SNIa.

Our second test of U Sco as a progenitor comes down to asking whether it is a neon nova.  The only way to recognize a neon nova with its neon-rich ejecta is by the high-excitation neon emission lines only visible during the late nebular phase of the eruption.  Neon has no readily visible lines in the optical spectrum, so the neon is usually recognized by the [Ne III] lines at 3869~\AA~and 3968~\AA~in the near ultraviolet.  The 3968~\AA~ line is always the weaker of the pair, and it can suffer from contamination by the H$\epsilon$ Balmer line, so the neon abundance is best given by the 3869~\AA~line alone.  When ultraviolet spectra are available, ``the [Ne V] (1575~\AA) and [Ne IV] (1602~\AA) emission lines have only been observed in nova eruptions from ONe WDs" (Darnley et al. 2017), while [Ne V] 3343~\AA~ and [Ne V] 3425~\AA~ are also pointers to neon novae.  The neon abundance can also be measured with X-ray spectra from late in the eruption.

Neon novae are those with a large over-abundance of neon in the ejecta, although we are not aware of any formal criterion.  Many papers identify specific novae as being neon nova, even though no abundance threshold is stated.  However, {\it all} nova explicitly identified as `neon novae' have neon $\geq$10$\times$ solar abundance of neon.  Further, {\it all} nova explicitly identified as being not neon have the calculated neon abundance $\leq$7$\times$ solar.  So it appear that the nova community has adopted a criterion that a nova with $\geq$10$\times$ solar is a neon nova.  The designation of `neon nova' only has utility for identifying eruptions where the neon is significantly more abundant than is possible from any cause other than dredge-up from an ONe WD.  The only place in the Universe to get large amounts neon is in the cores of middle-mass stars and their resultant ONe WDs.  In the context of U Sco, RNe, and CVs in general, the companion stars cannot provide any amount of neon much above solar composition.  And the nuclear burning cannot make any significant fraction of neon.  The threshold for no false alarms is going to be something like a factor of 10$\times$ solar, as there is no population of companion stars with a higher neon abundance.  Such a factor corresponds nicely to separate out the systems that researchers have explicitly labeled as `neon nova'.

A quick means to identify neon novae is to look at the line ratio of the [Ne III] 3869~\AA~line to the [O III] 5007~\AA~after extinction correction, $F_{3869}/F_{5007}$.  The spectra must be in the nebular phase, well after the transition, and even out into the early quiescence.  To correct for extinction, multiply the observed flux ratio times $10^{0.50*E(B-V)}$.  To set a threshold for the line ratio, we have taken the line ratios for 13 novae explicitly identified as `neon nova' in the literature as based on the traditional full radiative transfer elemental abundance analyses.  {\it All} of these neon novae have $F_{3869}/F_{5007}$$\geq$0.3.  Further, we have taken from the literature the 20 novae with full abundance analyses that are explicitly identified as not being neon nova, and {\it all} of these have $F_{3869}/F_{5007}$$\leq$0.25.  There is a perfect correspondence between the flux ratio with a threshold of 0.3 that recognizes neon novae.

There have been few published spectra of U Sco late in the tail of the eruption that extend down to at least 3700~\AA.  On Day $+$46 of the 2010 eruption, Mason (2011) recorded a spectrum extending down to 3000~\AA, with this showing a prominent [Ne III] 3869~\AA~line, as well as very prominent [Ne V] 3343~\AA~ and [Ne V] 3425~\AA~lines.  So already we know that U Sco is likely a neon nova.  The extinction corrected flux ratio $F_{3869}/F_{5007}$ is 0.40.  So, by the perfect criterion from the flux ratio, U Sco is easily a neon nova.

Mason (2011) also made a specific analysis of elemental abundances for U Sco.  She keyed off the [Ne/O] ratio, measured to be $+$1.69.  She collected the same ratio published for 10 known neon novae and found that they all have [Ne/O] greater than $+$0.45 and ranging up to $+$1.76.  For 12 known CO novae, all have [Ne/O] more negative than $+$0.11.  So the neon novae are clearly separated out by their [Ne/O] abundance analysis.  With this, U Sco looks to be a neon nova with particularly high neon abundance.

Diaz et al. (2012) used spectra from 3450--9000~\AA~on the SOAR 4.1-m telescope for Days $+$51 and $+$75.  They made extensive model calculations of fluxes and abundances over a wide range of conditions.  They conclude ``Both clumpy and spherically symmetric models require Neon overabundances of at least 10 times the solar values.''  Importantly, their analysis involved many lines from five elements and tested a wide range of temperatures and densities, so this result is robust.  (In particular, this Diaz result does not suffer from the possible ambiguity discussed in Mason 2013.)  That is, their full abundance analysis shows that U Sco is a neon nova.

Mason (2011) makes a new point about the ultraviolet line profiles, where all neon novae show prominent P-Cygni profiles after the iron curtain phase and before transition, whereas all the CO novae do not.  This would appear to be another strong indicator of whether the white dwarf is ONe or CO in composition.  Williams et al. (1981) report P-Cygni profiles during the same phase of the 1979 U Sco eruption.  So we have another strong bit of evidence saying the U Sco is a neon nova.  

Mason (2013) threw confusion into her prior analysis from Mason (2011) by publishing a corrigendum saying that the neon abundance can be made ambiguous for a special case.  That is, for the particular case of her original analysis method (involving just one flux ratio), the neon abundance is ``undetermined'' if the electron density is above $\sim$10$^6$ cm$^{-3}$.  However, all the cited cases show no ambiguity, in that the novae with high density are actually neon novae as determined with other analysis methods.  So we are not seeing evidence for any ambiguity.  Further, the case when the neon abundance might be ambiguous, requires a dubiously high density for the case of U Sco\footnote{For Day $+$104, an expansion velocity of 5700 km s$^{-1}$, and just 20 days for the ejection, the volume of the ejecta is $10^{47.5}$ cm$^3$.  This is a severe lower limit, because the observed velocity dispersion makes for a real volume roughly one order-of-magnitude larger.  We measured the 2010 eruption to have an ejecta mass of 26.5$\times$10$^{-6}$ $M_{\odot}$, which is the mass of $10^{52.5}$ hydrogen atoms.  If all of these atoms are ionized, the electron density would be $10^{5.0}$ cm$^{-3}$.  And this is an extreme upper limit.  With this case, the condition for ambiguity is missed by a factor of 10$\times$.  So the only way to reach the situation for ambiguity is for a very clumped shell, with a filling factor of $>$10$\times$.  To advocate for an ambiguous analysis, large variations in density are required, so already the applicability of the guessed temperature and density make for complex effects on the derived neon abundance.  That is, postulating the existence of high density regions has little impact on knowing the density for the regions that dominate the neon line emission. } on Days $+$46 to $+$104.  The potential ambiguity does not occur in the abundance analysis of Diaz et al. (2012), so their neon nova conclusion stands.  Mason (2013) goes on to reiterate the strong case for a neon nova due to the very prominent [Ne V] 3343~\AA~ and [Ne V] 3425~\AA~lines, plus the deep P-Cygni.  In all, we think that the confusion raised by the Mason corrigendum has not changed any conclusion.

What are the evidences as to whether U Sco is a neon nova?  First, the 3869~\AA~neon line is bright, with a flux ratio $F_{3869}/F_{5007}$=0.40, which gives an empirical proof that U Sco is a neon nova.  Second, Diaz et al. (2011) make a full abundance analysis to ``require Neon overabundances of at least 10 times the solar values.''  Third, Mason (2011) presents an abundance analysis that strongly concludes U Sco is a neon nova with some of the highest neon content known, although the later corrigendum weakly suggests that this one claim might have ambiguities for an unlikely extreme density.  Fourth, the [Ne V] 3343~\AA~ and [Ne V] 3425~\AA~lines are very prominent, pointing to a high neon abundance.  Fifth, the presence of prominent P-Cygni profiles in the ultraviolet spectra from a particular time just after peak light is a reliable indicator that U Sco is a neon nova.

So there we have it, much evidence demonstrates that U Sco is a neon nova.  The only way to get large bulk masses of neon into a nova ejecta is by dredge-up from the top layer of an underlying ONe white dwarf.  With U Sco having an ONe WD, it cannot possible become a normal Type Ia supernova, in any case.

\subsection{U Sco is {\it Not} a Supernova Progenitor}

We have run two tests for whether U Sco is a Type Ia supernova.  We find that U Sco has $M_{\rm ejecta}$ $\gg$ $M_{\rm accreted}$ and certainly cannot be a progenitor.  We find that U Sco is a neon nova with an ONe white dwarf and certainly cannot be a progenitor.  These results have no significant uncertainties and we cannot even imagine a way to impeach either conclusion.  U Sco will never evolve into a supernova.

\begin{acknowledgments}
We are thankful for the many professional-quality amateur astronomers contributing vast amounts of $UBVRI$ photometry for U Sco.  These workers are producing light curves for all galactic novae with coverage orders of magnitude beyond what is possible for any professional or satellite programs.  The AAVSO has provided many of the fundamental tools used in this 4-decade program.  These tools include the archiving of magnitude measures from top-quality observers worldwide, the comparison star magnitudes from APASS, and the coordination for the U Sco eruption campaigns.  The {\it TESS} data are publicly available at the Barbara A. Mikulski Archive for Space Telescopes.  
\end{acknowledgments}

%

\vspace{5mm}
\facilities{AAVSO, TESS}


{}



\begin{thebibliography}{99}

\bibitem[\protect\citeauthoryear{Ahnert}{1960}] {Ahnert 1960} 
Ahnert, P. 1960, Astr. Nach., 285 191
\bibitem[\protect\citeauthoryear{Anupama and Dewangan}{2000}] {Anupama and Dewangan 2000}
Anupama, G. C., \& Dewangan, G. C. 2000, AJ, 119, 1359
\bibitem[\protect\citeauthoryear{Chomiuk et al.}{2020}] {Chomiuk et al. 2020}
Chomiuk, L., Metzger, B. D., \& Shen, K. J. 2020, ARA\&A, 59, 391
\bibitem[\protect\citeauthoryear{Darnley et al.}{2017}] {Darnley et al. 2017} 
Darnley, M. J., Hounsell, R., Godon, P., et al.  2017, ApJ, 847, 35
\bibitem[\protect\citeauthoryear{Diaz et al.}{2012}] {Diaz et al. 2012} 
Diaz, M., Williams, R., Luna, G., Moraes, M., \& Takeda, L.  2012, Mem. S. A. It., 83, 758
\bibitem[\protect\citeauthoryear{Evans et al.}{2023}] {Evans et al. 2023} 
Evans, A., Banerjee, D. P. K., Woodward, C. E., et al.  2023, MNRAS, 522, 4841
\bibitem[\protect\citeauthoryear{Hachisu and Kato}{2019}] {Hachisu and Kato 2019} 
Hachisu, I., \& Kato, M. 2019, ApJS, 242, 18
\bibitem[\protect\citeauthoryear{Hillman and Kashi}{2021}] {Hillman and Kashi 2021}
Hillman, Y., \& Kashi, A. 2021, MNRAS, 501, 201
\bibitem[\protect\citeauthoryear{Jose and Hernanz}{1998}] {Jose and Hernanz 1998} 
Jos\'{e}, J. \& Hernanz, M. 1998, ApJ, 494, 680
\bibitem[\protect\citeauthoryear{Kato and Hachisu}{2012}] {Kato and Hachisu 2012} 
Kato, M., \& Hachisu, I.  2012, Bull. Astr. Soc. India, 40, 393
\bibitem[\protect\citeauthoryear{Livio}{2000}] {Livio 2000} 
Livio, M.  2000, in Type Ia Supernovae, Theory and Cosmology, eds J. C. Niemeyer and J. W. Truran.  Cambridge Univ. Press, p. 33
\bibitem[\protect\citeauthoryear{Livio et al.}{1991}] {Livio et al. 1991} 
Livio, M., Govarie, A., \& Ritter, H., 1991, A\&A, 246, 84
\bibitem[\protect\citeauthoryear{Maoz et al.}{2014}] {Maoz et al. 2014} 
Maoz, D., Mannucci, F., Nelemans, G., 2014, ARA\&A, 52, 107
\bibitem[\protect\citeauthoryear{Mason}{2011}] {Mason 2011} 
Mason, E.  2011, A\&A, 532, L11
\bibitem[\protect\citeauthoryear{Mason}{2013}] {Mason 2013} 
Mason, E.  2013, A\&A, 556, C2
\bibitem[\protect\citeauthoryear{Muraoka et al.}{2024}] {Muraoka et al. 2024} 
Muraoka, K., Kojiguchi, N., Ito, J., et al.  2024, PASJ, 76, 293
\bibitem[\protect\citeauthoryear{Muraoka et al.}{2025}] {Muraoka et al. 2025} 
Muraoka, K., Naoto, K., Ito, J., et al.  2025, PASJ, temp, 42
\bibitem[\protect\citeauthoryear{Pagnotta et al.}{2015}] {Pagnotta et al. 2015} 
Pagnotta, A., Schaefer, B. E., Clem, J. L., et al.  2015, ApJ, 811, 32     
\bibitem[\protect\citeauthoryear{Ricker et al.}{2015}] {Ricker et al. 2015}Ricker, G. R., Winn, J. N., Vanderspek, R., et al. 2015, Jour. Astron. Telescopes Instr. Systems, 1, 014003
\bibitem[\protect\citeauthoryear{Ruiter and Seitenzahl}{2025}] {Ruiter and Seitenzahl 2025} 
Ruiter, A. J., \& Seitenzahl, I. R. 2025, A\&ARev, 33, 1 
\bibitem[\protect\citeauthoryear{Rukeya et al.}{2017}] {Rukeya et al. 2017} 
Rukeya, R., Guoliang, L., Wang, Z., \& Zhu, C. 2017, PASP, 129, 74201
\bibitem[\protect\citeauthoryear{Salazar et al.}{2017}]{Salazar et al. 2017}		
Salazar, I. V., LeBleu, A., Schaefer, B. E., Landolt, A. U., \& Dvorak, S. 2017, MNRAS, 469, 4116
\bibitem[\protect\citeauthoryear{Schaefer}{1988}] {Schaefer 1988} 
Schaefer, B. E.  1988, ApJ, 327, 347	
\bibitem[\protect\citeauthoryear{Schaefer}{1988}] {Schaefer 1988} 
Schaefer, B. E.  1990, ApJ, 355, L39	
\bibitem[\protect\citeauthoryear{Schaefer}{2009}] {Schaefer 2009} 
Schaefer, B. E.  2009, ApJ, 697, 721	
\bibitem[\protect\citeauthoryear{Schaefer}{2010}] {Schaefer 2010} 
Schaefer, B. E.  2010, ApJS, 187, 275	
\bibitem[\protect\citeauthoryear{Schaefer}{2011}] {Schaefer 2011} 
Schaefer, B. E.  2011, ApJ, 742, 112     
\bibitem[\protect\citeauthoryear{Schaefer}{2020a}] {Schaefer 2020a} 
Schaefer, B. E. 2020a, MNRAS, 492, 3323		
\bibitem[\protect\citeauthoryear{Schaefer}{2020b}] {Schaefer 2020b} 
Schaefer, B. E.  2020b, MNRAS, 492, 3343      
\bibitem[\protect\citeauthoryear{Schaefer}{2021}] {Schaefer 2021} 
Schaefer, B. E. 2021, RNAAS, 5, 148			 
\bibitem[\protect\citeauthoryear{Schaefer}{2022a}] {Schaefer 2022a} 
Schaefer, B. E. 2022a, MNRAS, 516, 4497 		
\bibitem[\protect\citeauthoryear{Schaefer}{2022b}] {Schaefer 2022b} 
Schaefer, B. E. 2022b, MNRAS, 517, 3640 		
\bibitem[\protect\citeauthoryear{Schaefer}{2022c}] {Schaefer 2022c} 
Schaefer, B. E. 2022c, MNRAS, 517, 6150		
\bibitem[\protect\citeauthoryear{Schaefer}{2023a}] {Schaefer 2023a} 
Schaefer, B. E. 2023a, MNRAS, 524, 3146      
\bibitem[\protect\citeauthoryear{Schaefer}{2023b}] {Schaefer 2023b} 
Schaefer, B. E. 2023b, MNRAS, 525, 785		
\bibitem[\protect\citeauthoryear{Schaefer}{2024}] {Schaefer 2024} 
Schaefer, B. E. 2024, ApJ, 966, 155		
\bibitem[\protect\citeauthoryear{Schaefer}{2025a}] {Schaefer 2025a} 
Schaefer, B. E. 2025a, ApJ, 980, 156			 
\bibitem[\protect\citeauthoryear{Schaefer}{2025c}] {Schaefer 2025c} 
Schaefer, B. E. 2025c, ApJ, submitted			 
\bibitem[\protect\citeauthoryear{Schaefer et al.}{2013}] {Schaefer et al. 2013} 
Schaefer, B. E., Landolt, A. U., Linnolt, M., et al. 2013, ApJ, 773, 55     
\bibitem[\protect\citeauthoryear{Schaefer et al.}{1992}] {Schaefer et al. 1992} 
Schaefer, B. E., Landolt, A. U., Vogt, N., et al.  1992, ApJS, 81, 321		
\bibitem[\protect\citeauthoryear{Schaefer et al.}{2010a}] {Schaefer et al. 2010a} 
Schaefer, B. E., Pagnotta, A., Allen, B., et al. 2010a, ATel, 2452
\bibitem[\protect\citeauthoryear{Schaefer et al.}{2011}] {Schaefer et al. 2011} 
Schaefer, B. E., Pagnotta, A., LaCluyze, A. P., et al. 2011, ApJ, 742, 113	
\bibitem[\protect\citeauthoryear{Schaefer and Patterson}{1983}] {Schaefer  and Patterson 1983} 
Schaefer, B. E., \& Patterson J. O.  1983, ApJ, 268, 710
\bibitem[\protect\citeauthoryear{Schaefer and Ringwald}{1995}] {Schaefer and Ringwald 1995} 
Schaefer, B. E., \& Ringwald, F. A. 1995, ApJ, 447, L45	
\bibitem[\protect\citeauthoryear{Schaefer et al.}{2019}] {Schaefer et al. 2019} 
Schaefer, B. E., Boyd, D., Clayton, G. C., et al.  2019, MNRAS, 487, 1120		
\bibitem[\protect\citeauthoryear{Schlegel et al.}{2010}] {Schlegel et al. 2010} 
Schlegel, E. M., Schaefer, B., Pagnotta, A., et al. 2010, ATel, 2430
\bibitem[\protect\citeauthoryear{Shara et al.}{1986}] {Shara et al. 1986} 
Shara, M. M., Livio M., Moffatt A. F. J., \& Orio M., 1986, ApJ, 311, 163
\bibitem[\protect\citeauthoryear{Shara et al.}{2018}] {Shara et al. 2018}
Shara, M. M., Prialnik, D., Hillman, Y., \& Kovetz, A. 2018, ApJ, 860, 110
\bibitem[\protect\citeauthoryear{Shen and Bildsten}{2009}] {Shen and Bildsten 2009} 
Shen, K. J., \& Bildsten, L. 2009, ApJ, 699, 1365
\bibitem[\protect\citeauthoryear{Shen and Quataert}{2022}] {Shen and Quataert 2022} 
Shen, K. J., \& Quataert, E. 2022, ApJ, 938, 31
\bibitem[\protect\citeauthoryear{Sparks and Sion}{2021}] {Sparks and Sion 2021} 
Sparks, W. M., \& Sion, E. M. 2021, ApJ, 914, 5
\bibitem[\protect\citeauthoryear{Starrfield et al.}{2024}] {Starrfield et al. 2024} 
Starrfield, S., Bose, M., Iliadis, C., et al. 2024, ApJ, 962, 191
\bibitem[\protect\citeauthoryear{Starrfield et al.}{2025}] {Starrfield et al. 2025} 
Starrfield, S., Bose, M., Woodward, C. E., et al. 2025, ApJ, 982, 89
\bibitem[\protect\citeauthoryear{Starrfield et al.}{1986}] {Starrfield et al. 1986} 
Starrfield, S., Sparks, W. M., \& Truran, J. W. 1986, ApJ, 303, L5		
\bibitem[\protect\citeauthoryear{Strope, Schaefer, and Henden}{2010}] {Strope, Schaefer, and Henden 2010}Strope, R. J., Schaefer, B. E., \& Henden, A. A.  2010, AJ, 140, 34	
\bibitem[\protect\citeauthoryear{Thoroughgood et al.}{2001}] {Thoroughgood et al. 2001} 
Thoroughgood, T. D., Dhillon, V. S., Littlefair, S. P., Marsh, T. R., \& Smith, D. A.  2001, MNRAS, 327, 1323
\bibitem[\protect\citeauthoryear{Truran and Livio}{1986}] {Truran and Livio 1986} 
Truran, J. W., \& Livio, M. 1986, ApJ, 308, 721
\bibitem[\protect\citeauthoryear{Walder et al.}{2008}] {Walder et al. 2008} 
Walder, R., Folini, D., \& Shore, S. N. 2008, A\&A, 484, L9
\bibitem[\protect\citeauthoryear{Wang and Han}{2012}] {Wang and Han 2012} 
Wang, B., \& Han, Z.  2012, New Astron. Rev., 56, 122
\bibitem[\protect\citeauthoryear{Williams et al.}{1981}] {Williams et al. 1981} 
Williams, R. E., Sparks, W. M., Gallagher, J. S., et al.  1981, ApJ, 251, 221
\bibitem[\protect\citeauthoryear{Yaron et al.}{2005}] {Yaron et al. 2005} 
Yaron, O., Prialnik, D., Shara, M. M., \& Kovetz A.  2005, ApJ, 623, 398
\bibitem[\protect\citeauthoryear{Zwitter and Munari}{2000}] {Zwitter and Munari 2000} 
Zwitter, T., \& Munari, U. 2000, New Astron. Rev., 44, 67


\end{thebibliography}
\end{document}